\documentclass[aps, prd, notitlepage, letterpaper, 11pt, amsmath, amssymb, amsfonts, floatfix, nofootinbib, superscriptaddress]{revtex4-1}

\usepackage[utf8]{inputenc}
\usepackage[english]{babel}
\usepackage{microtype}
\usepackage{graphicx} 
\usepackage{dcolumn}
\usepackage{slashed}
\usepackage{lineno}
\usepackage{color}
\usepackage{amsmath}
\usepackage{amssymb}
\usepackage{caption}
\usepackage{subcaption}
\usepackage{xcolor}
\usepackage{cancel}
\usepackage{framed}
\usepackage{comment}
\usepackage{mathtools}
\usepackage{hyperref}
\usepackage{indentfirst}
\usepackage{multirow}

\usepackage{ulem}

\definecolor{added}{rgb}{0,0,1}
\definecolor{deleted}{rgb}{1,0,0}

\newcommand{\overbar}[1]{\mkern 1.5mu\overline{\mkern-1.5mu#1\mkern-1.5mu}\mkern 1.5mu}

\makeatletter
\newcommand*{\toccontents}{\@starttoc{toc}}
\makeatother

\allowdisplaybreaks

\begin{document}

\title{Rare Top Decays as Probes of Flavorful Higgs Bosons}

\def\ucsc{Santa Cruz Institute for Particle Physics, University of California, Santa Cruz, CA 95064, USA}

\author{Wolfgang~Altmannshofer} \email[]{waltmann@ucsc.edu}
\affiliation{\ucsc}

\author{Brian~Maddock} \email[]{bmaddock@ucsc.edu}
\affiliation{\ucsc}

\author{Douglas~Tuckler} \email[]{dtuckler@ucsc.edu}
\affiliation{\ucsc}

\begin{abstract}
We study a version of Two Higgs Doublet Models with non-standard flavor violation in the up quark sector.
We find branching ratios for the rare top decays $t\to hc$ and $t\to hu$ that are within reach of current and future colliders, while other flavor constraints from rare $B$ decays and neutral $D$ meson mixing, as well as constraints from Higgs signal strength measurements remain under control. The most prominent collider signature of the considered setup is $pp \to t H \to tt \bar c$, providing continued motivation to search for same-sign tops at the LHC as well as a simple framework to interpret these searches. As a byproduct of our study, we provide updated SM predictions for the rare top decays BR$(t \to h c)_\text{SM} = (4.19^{+1.08}_{-0.80} \pm 0.16) \times 10^{-15}$ and BR$(t \to h u)_\text{SM} = (3.66^{+0.94}_{-0.70} \pm 0.67) \times 10^{-17}$ with the main uncertainties coming from higher order QCD and CKM matrix elements.
\end{abstract}

\maketitle

\section{Introduction}

Flavor-changing neutral current (FCNC) decays of the top quark appear at one-loop in the Standard Model (SM) and are strongly suppressed by the Glashow-Iliopoulos-Maiani (GIM) mechanism~\cite{Glashow:1970gm} and the small mixing of the third generation quarks with the first and second generations. In particular, the branching ratios of the rare decays $t \rightarrow hc$ and $t \rightarrow hu$ in the SM are predicted to be $\mathcal{O}(10^{-15})$ and $\mathcal{O}(10^{-17})$~\cite{Mele:1998ag,AguilarSaavedra:2004wm}, respectively, which renders these processes unobservable in the foreseeable future~\cite{Aaboud:2018oqm, Sirunyan:2017uae, ATLAS:url, deBlas:2018mhx, Cerri:2018ypt, Mangano:2018mur}. Observation of such processes at current or planned colliders would be a clear signal of physics beyond the SM.

The mass of the top quark is close to the electroweak scale and large compared to the masses of the rest of the quarks which suggests that it is intimately connected with the mechanism of electroweak symmetry breaking (EWSB). By now it has been well established that the main source of mass generation for the weak gauge bosons and the third generation fermions is the vacuum expectation value (vev) of the 125 GeV Higgs boson discovered at the Large Hadron Collider (LHC)~\cite{Sirunyan:2017khh, Aaboud:2018pen, Sirunyan:2018kst, Aaboud:2018zhk, Sirunyan:2018hoz, Aaboud:2018urx}. However, much less is known about the origin of mass for the first and second generations of fermions. Although constraints have been placed on the 125 GeV Higgs coupling to the muon~\cite{Sirunyan:2018hbu,Aaboud:2017ojs}, direct measurements of its couplings to the first and second generation fermions are experimentally difficult, and it is presently unknown if the 125 GeV Higgs gives mass to the light fermions. In addition, the masses of the fermions as well as the Cabibbo-Kobayashi-Maskawa (CKM) quark mixing matrix exhibit a hierarchical structure that is not \textit{a priori} explained by the SM -- this is known as the SM flavor puzzle. 

The SM flavor puzzle can be partially addressed by introducing an additional source of EWSB that is responsible for generating mass for the first and second generation fermions, as proposed in~\cite{Altmannshofer:2015esa} (see also~\cite{Ghosh:2015gpa,Botella:2016krk,Das:1995df,Blechman:2010cs}). Arguably the simplest realization of this scenario is a Two Higgs Doublet Model (2HDM), in which one Higgs doublet (approximately identified as the 125 GeV Higgs boson) couples mainly to the third generation, while the second Higgs doublet couples mainly to the first and second generations. The observed pattern of quark masses and mixing can be obtained by asserting suitable textures for the quark mass matrices, leading to a ``flavorful'' Two Higgs Doublet Model (F2HDM). The Yukawa structure of the F2HDM can be for example generated via the flavor-locking mechanism~\cite{Knapen:2015hia,Altmannshofer:2017uvs}, and the implied collider phenomenology has been discussed in detail in~\cite{Altmannshofer:2016zrn}. Variations of the original F2HDM with alternative flavor structures were explored in~\cite{Altmannshofer:2018bch} (for additional recent work on extended Higgs sectors with non-standard flavor structures see~\cite{Bauer:2015fxa, Chiang:2015cba, Botella:2015hoa, Bauer:2015kzy, Buschmann:2016uzg, Sher:2016rhh, Primulando:2016eod, Alves:2017xmk, Gori:2017tvg, Crivellin:2017upt, Kohda:2017fkn, Dery:2017axi, Banerjee:2018fsx, Diaz-Cruz:2019emo, Ferreira:2019aps}).

In this work, we will explore the effects of the F2HDM on rare top quark decays $t \to h u$ and $t \to h c$, where $h$ is the SM-like Higgs. 
In previous studies, we considered setups where the CKM matrix originates in the down quark sector, i.e. the CKM matrix is largely given by the matrix that diagonalizes the down quark Yukawa couplings. Such a setup is a natural choice given the hierarchies in the down quark masses and the CKM matrix elements are comparable, $V_{us} \sim m_d/m_s$, $V_{cb} \sim m_s/m_b$, $V_{ub} \sim m_d/m_b$. Such setups can lead to enhanced flavor violating couplings of the Higgs bosons to down type quarks, resulting in potentially interesting effects in $B$ meson oscillations and rare $B$ decays. Rare top decays, however, tend to be strongly suppressed. In this work we will instead explore setups in which the CKM matrix is generated in the up quark sector, which can lead to enhanced tree level flavor violating couplings of the Higgs bosons to up type quarks. These couplings can produce branching ratios for the rare top quark decays $t\to hu$ and $t\to hc $ that are orders of magnitudes greater than the SM predictions, and can be within reach of current and future colliders.

Because the mass hierarchies in the up quark sector are rather different than those in the down quark sector, the flavor-locking mechanism \cite{Knapen:2015hia,Altmannshofer:2017uvs} is not suitable for generating the CKM matrix in the up sector.
Thus, we will consider a scenario where the required up Yukawa textures are dynamically generated by a Froggatt-Nielsen (FN) type mechanism \cite{Froggatt:1978nt}.

The paper is structured as follows. In Sec.~\ref{sec:F2HDM} we discuss F2HDMs with a CKM matrix that originates in the up sector and identify a setup that dynamically generates the required flavor structure of the up Yukawa using the FN mechanism. In Sec.~\ref{sec:RareB} we consider the stringent constraints on the model from the rare decay $b\rightarrow s \gamma$. In Sec.~\ref{sec:RareTop} we first update the SM predictions for the branching ratios of the rare decays $t \to hu$ and $t\to hc$. We then study how these decays are affected by the tree-level flavor-changing Higgs couplings $htc$ and $htu$ and compare the F2HDM predictions for the $t \to hu$ and $t\to hc$ branching ratios with existing and expected experimental sensitivities. In Sec.~\ref{sec:signatures} we discuss related effects of the model on neutral $D$ meson mixing and the collider phenomenology of the heavy Higgs bosons, identifying features that are different from the down type F2HDMs studied in~\cite{Altmannshofer:2016zrn,Altmannshofer:2018bch}. We conclude in Sec.~\ref{sec:conclusion}.

\section{Flavorful 2HDMs with Up Sector CKM}\label{sec:F2HDM}

In a generic 2HDM the interactions of the two Higgs doublets, $\Phi$ and $\Phi^\prime$ (with vevs $v$ and $v^\prime$), with the SM quarks and leptons are described by the Yukawa Lagrangian
\begin{align}
\label{eq:yuk}
-\mathcal{L}_\text{Yuk} = \sum_{i,j} \bigg( \lambda^u_{ij}&(\overbar{Q}_i u_j)  \tilde{\Phi} +  \lambda^d_{ij}(\overbar{Q}_i d_j) \Phi +  \lambda^e_{ij}(\bar{\ell}_i e_j) \Phi \nonumber \\ 
& +\lambda^{\prime u}_{ij}(\overbar{Q}_i u_j) \tilde{\Phi}^\prime +  \lambda^{\prime d}_{ij}(\overbar{Q}_i d_j) \Phi^\prime +  \lambda^{\prime e}_{ij}(\bar{\ell}_i e_j) \Phi^\prime \bigg) +\text{ h.c.} ~,
\end{align}
where $\tilde{\Phi}^{(\prime)} = i \sigma_2 (\Phi^{(\prime)})^*$, $q_i, \ell_i$ are the three generations of the left-handed quark and lepton doublets, and  $u_i, d_i, e_i$ are the three generations of right-handed up quark, down quark, and charged lepton singlets. Generically, the Yukawa matrices $\lambda^{ (\prime) u,d,\ell}$ will contain off-diagonal entries that generate flavor-violating processes at tree-level. In order to avoid tensions with low energy flavor constraints, one often imposes a discrete $\mathbb{Z}_2$ symmetry on the Higgs and quark fields such that the couplings of the Higgs bosons are flavor diagonal, leading to the well studied 2HDMs with natural flavor conservation: type~I, type~II, flipped, and lepton-specific~\cite{Glashow:1976nt}.

A `flavorful' 2HDM, as introduced in~\cite{Altmannshofer:2015esa,Altmannshofer:2018bch}, does not impose these discrete symmetries and instead assumes that one set of the Yukawa couplings are rank 1, preserving a $U(2)^5$ symmetry acting on the first two generations that is minimally broken by the second set of Yukawa couplings. In this way, flavor transitions between the first and second generations are protected and appear only at second order as an effective transition. In~\cite{Altmannshofer:2018bch} four such models we identified, that, in analogy to the models with $\mathbb{Z}_2$ symmetry, were denoted by type~IB, type~IIB, flipped~B, and lepton-specific~B 2HDMs. In tab.~\ref{tab:typeB} we summarize which Higgs boson is primarily responsible for generating the masses of each fermion. 

In addition to reproducing the observed quark masses, the F2HDMs also need to accommodate the CKM quark mixing matrix.
The CKM matrix arises from the mismatch of rotations of left-handed up and down type quarks when rotating into the quark mass eigenstate basis. The CKM matrix can originate dominantly from the rotations in the up quark sector or from those in the down quark sector. In previous studies~\cite{Altmannshofer:2015esa,Altmannshofer:2016zrn,Altmannshofer:2017uvs,Altmannshofer:2018bch} the CKM matrix was generated in the down quark sector. In this work we will instead generate the CKM matrix in the up quark sector. 

\bgroup
\def\arraystretch{1.5}
\setlength\tabcolsep{6pt}
\begin{table}[tb]
\begin{center}
\begin{tabular}{l cccccc}
\hline\hline
Model & $u, c$ & $t$ & $d,s$ & $b$ & $e,\mu$ & $\tau$ \\ 
\hline
Type 1B & $\Phi'$& $\Phi$&$\Phi'$& $\Phi$& $\Phi'$&$\Phi$ \\
Type 2B& $\Phi'$& $\Phi$& $\Phi$& $\Phi'$& $\Phi$& $\Phi'$\\
Flipped B& $\Phi'$& $\Phi$&$\Phi$& $\Phi'$& $\Phi'$&$\Phi$ \\
Lepton-Specific B &$ \Phi'$& $\Phi$& $\Phi'$ &$ \Phi$& $\Phi$& $\Phi'$ \\
\hline\hline
\end{tabular}
\end{center}
\caption{Dominant source of mass for the SM fermions in F2HDMs.}
\label{tab:typeB}
\end{table}
\egroup

Since the hierarchies in the up quark masses are different than those in the CKM matrix, the flavor-locking mechanism is not suitable for generating appropriate Yukawa textures for an ``up type'' F2HDM. We will therefore use the Froggatt-Nielsen mechanism, which explains the hierarchy of quark masses and mixing by introducing an abelian flavor symmetry -- which we will denote by $U(1)_\text{FN}$ -- that distinguishes different fermion flavors. The flavor symmetry is broken by the vev of a SM-singlet scalar field, $S$, that carries a $U(1)_\text{FN}$ charge $Q_S = +1$. This breaking is communicated to the SM fermions by higher dimensional operators leading to Yukawa couplings that are suppressed by powers of a small symmetry-breaking parameter $\epsilon = \langle S \rangle / \Lambda_S$, where $\Lambda_S \gg v,v^\prime$ is the scale associated with the breaking of $U(1)_\text{FN}$. In the resulting effective theory, the Yukawa Lagrangian is given by \footnote{Here we only describe the up quark sector, but an analogous discussion applies for down quarks and leptons as well.}
\begin{equation}
-\mathcal{L}^\text{eff}_\text{Yuk} \supset \sum_{i,j} \Bigg[ \bigg( \frac{\langle S \rangle}{\Lambda_S} \bigg)^{|x^u_{ij}|}(\overbar{Q}_i u_j)  \tilde{\Phi} +\bigg(\frac{\langle S \rangle}{\Lambda_S} \bigg)^{|x^{\prime u}_{ij}|}(\overbar{Q}_i u_j) \tilde{\Phi}^\prime \Bigg],
\end{equation}
where the powers $x^{(\prime)u}_{ij}$ are determined from the charge assignments of the Higgs and quark fields under $U(1)_\text{FN}$, and we have omitted model dependent prefactors of $O(1)$. In terms of the parameter $\epsilon \approx 0.22$ we aim for the following relations for the quark masses and the CKM matrix elements
\begin{eqnarray} \label{eq:FNpowers}
&& \frac{m_u}{v_\text{w}} \sim \epsilon^8~,~~ \frac{m_c}{v_\text{w}}\sim \epsilon^3 ~,~~ \frac{m_t}{v_\text{w}}\sim \epsilon^0 ~,~~ \frac{m_d}{v_\text{w}} \sim \epsilon^7~,~~ \frac{m_s}{v_\text{w}}\sim \epsilon^5 ~,~~ \frac{m_b}{v_\text{w}}\sim \epsilon^3 ~, \nonumber \\
&& |V_{us}| \sim \lambda_c \sim \epsilon ~,~~ |V_{ub}| \sim \lambda_c^3 \sim \epsilon^3 ~,~~|V_{cb}| \sim \lambda_c^2 \sim \epsilon^2 ~,
\end{eqnarray}
with the electroweak breaking vev in the SM $v_\text{w} = \sqrt{v^2 + v^{\prime 2}} \simeq 246$~GeV and the Cabibbo angle $\lambda_c \simeq 0.22$.

In order to obtain the rank 1 structure of the Yukawa couplings of $\Phi$ required by the F2HDM scenario, we introduce an additional $U(1)^\prime$ symmetry.
A rank 1 Yukawa coupling $\lambda_u$ and simultaneous generation of the CKM matrix by $\lambda^\prime_u$ is possible by charging either the left-handed quark doublet $Q_3$ or the right-handed top $U_3$ under the additional symmetry. 2HDMs with a right-handed top that is singled out by a symmetry have been discussed e.g. in~\cite{Chiang:2015cba,Gori:2017tvg}. Here we follow the second option and set the $U(1)^\prime$ charges $Q^\prime_\Phi = - Q^\prime_{Q_3} = +1$, while leaving the right-handed top uncharged. We will see that this leads to highly predictive scenarios.

The remaining charge assignments depend on the type of F2HDM under consideration as well as on the value of the parameter $\tan\beta = v/v^\prime$ that can provide part of the fermion mass hierarchies.
We restrict the following discussion to the quark sector. The extension to charged leptons is straight forward.

\bgroup
\def\arraystretch{1.5}
\setlength\tabcolsep{5pt}
\begin{table}[tb]
\begin{center}
\begin{tabular}{c c ccc ccccccccc}
\hline\hline
& & $S$ & $\Phi$ &$\Phi^\prime$~&~ $\overbar{Q}_1$~&~$\overbar{Q}_2$~&~$\overbar{Q}_3$~&~$u_1$~&~$u_2$~&~$u_3$~&~$d_1$~&~$d_2$~&~$d_3$\\
\hline
$\tan\beta \sim 1/\epsilon$ &$U(1)_\text{FN}$~~& 1 & ~0 ~& 0 ~&~ 2~&~1~&~0~&~5 or -9~&~1~&~0~&~ 4 or -8 ~&~ 3 ~&~ $\pm3$ \\
$\tan\beta \sim 1/\epsilon^2$ &$U(1)_\text{FN}$~~& 1 & ~0 ~& 0 ~&~ 1~&~0~&~0~&~5 or -7~&~1~&~0~&~ 4 or -6 ~&~ 3 ~&~ $\pm3$ \\
\hline
& $U(1)^\prime$~~& 0 &1 ~& 0 ~&~ 0~&~0~&~1~&~0~&~0~&~0~&~0~&~0~&~-2\\
\hline\hline
\end{tabular}
\end{center}
\caption{Charges of the Froggatt-Nielsen scalar $S$, the two Higgs doublets $\Phi$ and $\Phi^\prime$ and quark fields under the $U(1)_\text{FN}$ and $U(1)^\prime$ symmetries in the type~IB and lepton-specific~B models for the two choices of $\tan\beta \sim 1/\epsilon$ and $\tan\beta \sim 1/\epsilon^2$.} \label{tab:charges}
\end{table}
\egroup

\paragraph{Type~IB and Lepton-Specific~B:} In these types, the coupling of $\Phi$ to both up type and down type quarks are rank 1. 
Given our choice of $U(1)^\prime$ charges discussed above, the charge of the right-handed bottom quark is required to be $Q^\prime_{d_3} = -2$. While all other quarks remain uncharged under the $U(1)^\prime$. For a given value of $\tan\beta$, the scaling in~(\ref{eq:FNpowers}) fixes all $U(1)_\text{FN}$ Froggatt-Nielsen charges up a few discrete choices. In Tab.~\ref{tab:charges} we show all inequivalent charge assignments in the cases $\tan\beta \sim 1/\epsilon \sim 5$ and $\tan\beta \sim 1/\epsilon^2 \sim 25$. The charge assignments lead to the following structure for the Yukawa couplings
\begin{subequations}
 \begin{align} \label{eq:lambda_u}
 v \lambda_u \sim v_\text{w} \begin{pmatrix}
~0~~&~~0~~&~~0~~\\
~0~~&~~0~~&~~0~~\\
 ~ \epsilon^{|a|}  ~~&~~  \epsilon^1 ~~&~~1~~
\end{pmatrix}, ~~~~~
v^\prime \lambda_u' \sim v_\text{w} \begin{pmatrix}
 ~\epsilon^8 ~~&~~ \epsilon^4 ~~&~~ \epsilon^3~~ \\
~ \epsilon^{|b|} ~~&~~ \epsilon^3 ~~&~~ \epsilon^2 ~~\\
~ 0 ~~&~~0 ~~&~~ 0~~
\end{pmatrix}, 
\\ \label{eq:lambda_d}
v \lambda_d \sim v_\text{w} \begin{pmatrix}
 ~0~~&~~0~~&~~0~~\\
 ~0~~&~~0~~&~~0~~\\
 ~0~~&~~0~~&~~\epsilon^3~~
\end{pmatrix}, ~~~~~
v^\prime \lambda_d' \sim v_\text{w} \begin{pmatrix}
 ~\epsilon^7 ~~&~~ \epsilon^6 ~~&~~ 0~~ \\
 ~\epsilon^{|c|} ~~&~~ \epsilon^5 ~~&~~0~~ \\
 ~0 ~~&~~0 ~~&~~ 0~~
\end{pmatrix},
\end{align}
\end{subequations}
with the powers $|a| = 5$ or $7$ or $9$, $|b| = 7$ or $9$, and $|c| = 6$ or $8$, depending on the charge assignments and $\tan\beta$. 
It is easy to check that the diagonalization of the quark masses that are induced by these Yukawa couplings leads to a CKM matrix with the right texture that is indeed dominantly generated from the up quark rotation.  
Interestingly, the powers $|a|$, $|b|$, and $|c|$ are not observable in the IR. 
More importantly, in the quark mass eigenstate basis, the flavor structure of all couplings of the Higgs bosons are entirely determined by the known quark masses and CKM elements. 
The couplings of the physical Higgs mass eigenstates $h,H,A,H^\pm$ to the quarks can be parameterized by
\begin{eqnarray}
  -\mathcal{L}_\text{Yuk} &\supset& \sum_{i,j}
   (\bar{d}_i P_R d_j) \Big(h(Y_h^d)_{ij}+H(Y^d_H)_{ij}-iA(Y_A^d)_{ij} \Big) + \text{h.c.}  \nonumber \\
  && + \sum_{i,j}(\bar{u}_i P_R u_j) \Big(h(Y_h^u)_{ij}+H(Y^u_H)_{ij}+iA(Y_A^u)_{ij} \Big) + \text{h.c.} \\
  && + \sqrt{2} \sum_{i,j} \Big( (\bar{d_i}P_R u_j)H^-(Y_{\pm}^u)_{ij}-(\bar{u_i}P_R d_j)H^+(Y^d_{\pm})_{ij} \Big) + \text{h.c.} ~. \nonumber
\end{eqnarray}
For all charge assignments we find for the up quark couplings 
\begin{subequations}
\begin{eqnarray}
v_\text{w} Y_h^u &=& \frac{c_\alpha}{s_\beta} \begin{pmatrix}  0 & 0 & 0 \\ 0 & 0 & 0 \\ 0 & 0 & m_t \end{pmatrix} -\frac{s_\alpha}{c_\beta} \begin{pmatrix}  m_u & 0 & 0 \\ 0 & m_c & 0 \\ 0 & 0 & 0 \end{pmatrix}
+ \frac{c_{\beta-\alpha}}{s_\beta c_\beta} \hat M_u ~, \\
v_\text{w} Y_H^u &=& \frac{1}{t_\beta} \frac{s_\alpha}{c_\beta} \begin{pmatrix}  0 & 0 & 0 \\ 0 & 0 & 0 \\ 0 & 0 & m_t \end{pmatrix} + t_\beta \frac{c_\alpha}{s_\beta} \begin{pmatrix}  m_u & 0 & 0 \\ 0 & m_c & 0 \\ 0 & 0 & 0 \end{pmatrix}
- \frac{s_{\beta - \alpha}}{s_\beta c_\beta} \hat M_u ~, \\
v_\text{w} Y_A^u &=& -\frac{1}{t_\beta} \begin{pmatrix}  0 & 0 & 0 \\ 0 & 0 & 0 \\ 0 & 0 & m_t \end{pmatrix} + t_\beta \begin{pmatrix}  m_u & 0 & 0 \\ 0 & m_c & 0 \\ 0 & 0 & 0 \end{pmatrix}
- \frac{1}{s_\beta c_\beta} \hat M_u ~, \\
v_\text{w} Y_\pm^u &=& -\frac{1}{t_\beta} \begin{pmatrix}  0 & 0 & 0 \\ 0 & 0 & 0 \\  m_u V_{ub}^* & m_c V_{cb}^* & m_t V_{tb}^* \end{pmatrix} + t_\beta \begin{pmatrix}  m_u V_{ud}^* & m_c V_{cd}^* & m_t V_{td}^* \\ m_u V_{us}^* & m_c V_{cs}^* & m_t V_{ts}^* \\ 0 & 0 & 0 \end{pmatrix},
\end{eqnarray}
\end{subequations}
and for the down quark couplings
\begin{subequations}
\begin{eqnarray}
v_\text{w} Y_h^d &=& \frac{c_\alpha}{s_\beta} \begin{pmatrix}  0 & 0 & 0 \\ 0 & 0 & 0 \\ 0 & 0 & m_b \end{pmatrix} -\frac{s_\alpha}{c_\beta} \begin{pmatrix}  m_d & 0 & 0 \\ 0 & m_s & 0 \\ 0 & 0 & 0 \end{pmatrix}, \\
v_\text{w} Y_H^d &=& \frac{1}{t_\beta} \frac{s_\alpha}{c_\beta} \begin{pmatrix}  0 & 0 & 0 \\ 0 & 0 & 0 \\ 0 & 0 & m_b \end{pmatrix} + t_\beta \frac{c_\alpha}{s_\beta} \begin{pmatrix}  m_d & 0 & 0 \\ 0 & m_s & 0 \\ 0 & 0 & 0 \end{pmatrix}, \\
v_\text{w} Y_A^d &=& -\frac{1}{t_\beta} \begin{pmatrix}  0 & 0 & 0 \\ 0 & 0 & 0 \\ 0 & 0 & m_b \end{pmatrix} + t_\beta \begin{pmatrix}  m_d & 0 & 0 \\ 0 & m_s & 0 \\ 0 & 0 & 0 \end{pmatrix}, \\
v_\text{w} Y_\pm^d &=& -\frac{1}{t_\beta} \begin{pmatrix}  0 & 0 & V_{ub} m_b  \\ 0 & 0 & V_{cb} m_b \\ 0 & 0 & V_{tb} m_b \end{pmatrix} + t_\beta \begin{pmatrix}  V_{ud} m_d & V_{us} m_s & 0 \\ V_{cd} m_d & V_{cs} m_s & 0 \\ V_{td} m_d & V_{ts} m_s & 0 \end{pmatrix},
\end{eqnarray}
\end{subequations}
The angle $\alpha$ in the above expressions parameterizes the mixing between the neutral scalar Higgs bosons $h$ and $H$.
The mass matrix $\hat M_u$ that enters the up quark couplings is given by
\begin{equation}
\hat M_u = \begin{pmatrix} m_u |V_{ub}|^2 & m_c V_{ub} V_{cb}^* &  m_t V_{ub} V_{tb}^* \\ m_u V_{cb} V_{ub}^* & m_c |V_{cb}|^2 &  m_t V_{cb} V_{tb}^* \\ m_u V_{tb} V_{ub}^* & m_c V_{tb} V_{cb}^* & -m_t (|V_{cb}|^2+|V_{ub}|^2) \end{pmatrix}.
\end{equation}
The proof that the flavor structure of the Higgs couplings in the type~IB and lepton-specific~B models is indeed entirely determined by known quark masses and CKM elements is given in the appendix~\ref{mass_eigenstates}.
Note that the neutral Higgs couplings are flavor diagonal in the down sector. Therefore, there are no tree level contributions to e.g. $B$ and $K$ meson oscillations and rare $B$ meson decays. In the up sector, the neutral Higgs couplings are flavor violating but the amount of flavor violation is controlled by the CKM matrix. Remarkably, the only free parameters in the couplings are $\tan\beta$ and the Higgs mixing angle $\alpha$, making the type~IB and lepton-specific~B models with up-sector CKM highly predictive.

\paragraph{Type~IIB and Flipped~B:} In these types, the coupling of $\Phi$ to the up type quarks and the coupling of $\Phi^\prime$ to the down type quarks are rank 1. We find that with our $U(1)^\prime \times U(1)_\text{FN}$ setup, it is not possible to construct Yukawa matrices for the down type quarks that exactly mirror the couplings in Eq.~\eqref{eq:lambda_d}, but with the role of $\lambda^d$ and $\lambda^{\prime d}$ exchanged. 

\bgroup
\def\arraystretch{1.5}
\setlength\tabcolsep{5pt}
\begin{table}[tb]
\begin{center}
\begin{tabular}{c c ccc ccccccccc}
\hline\hline
& & $S$ & $\Phi$ &$\Phi^\prime$& $\overbar{Q}_1$&$\overbar{Q}_2$&$\overbar{Q}_3$&$u_1$&$u_2$&$u_3$&$d_1$&$d_2$&$d_3$\\
\hline
$\tan\beta \sim 1/\epsilon$ &$U(1)_\text{FN}$& 0 & 0 & 0&0&+1&+2&-7&-3&  -2 & -7 & -6&-4 \\
\hline
&$U(1)^\prime$& 0 & +1 & -1&+1&+1&-1&0&0&0&0&0&0\\
\hline\hline
\end{tabular}
\end{center}
\caption{Example charges of the Froggatt-Nielsen scalar $S$, the two Higgs doublets $\Phi$ and $\Phi^\prime$ and quark fields under the $U(1)_\text{FN}$ and $U(1)^\prime$ symmetries in the type~IIB and lepton-specific~B models for $\tan\beta \sim 1/\epsilon$.\label{tab:charges_2}} 
\end{table}
\egroup

However, we find that the $\lambda^{\prime d}$ couplings can still be made rank 1, with a consistent flavor structure as long as $\tan\beta \sim 1/\epsilon \sim 5$. In contrast to the type~IB and lepton-specific~B setups discussed above, we find that $\lambda^d$ and $\lambda^{\prime d}$ necessarily contain mixing between the third and the first two generations. One example set of charges is given in Tab.~\ref{tab:charges_2} which leads to 
\begin{subequations}
 \begin{align}
 v \lambda_u \sim v_\text{w} \begin{pmatrix}
~0~~&~~0~~&~~0~~\\
~0~~&~~0~~&~~0~~\\
 ~ \epsilon^5  ~~&~~  \epsilon^1 ~~&~~\epsilon^0~~
\end{pmatrix}, ~~~~~
v^\prime \lambda_u' \sim v_\text{w} \begin{pmatrix}
 ~\epsilon^8 ~~&~~ \epsilon^4 ~~&~~ \epsilon^3~~ \\
~ \epsilon^7 ~~&~~ \epsilon^3 ~~&~~ \epsilon^2 ~~\\
~ 0 ~~&~~0 ~~&~~ 0~~
\end{pmatrix}, 
\\
v \lambda_d \sim v_\text{w} \begin{pmatrix}
 ~\epsilon^7~~&~~\epsilon^6~~&~~\epsilon^4~~\\
 ~\epsilon^6~~&~~\epsilon^5~~&~~\epsilon^3~~\\
 ~0~~&~~0~~&~~0~~
\end{pmatrix}, ~~~~~
v^\prime \lambda_d' \sim v_\text{w} \begin{pmatrix}
 ~0 ~~&~~ 0 ~~&~~ 0~~ \\
 ~0 ~~&~~ 0 ~~&~~ 0~~ \\
 ~\epsilon^6 ~~&~~ \epsilon^5 ~~&~~ \epsilon^3~~
\end{pmatrix}.
\end{align}
\end{subequations}
The more generic structure of the down quark Yukawas implies that the CKM matrix is partly generated also from the rotations in the down sector.
Correspondingly, in the type~IIB and flipped~B models only the generic scaling of the couplings of the physical Higgs bosons can be predicted. The precise values of the physical Higgs couplings depend on unknown $O(1)$ parameters. 

As we will see in Sec.~\ref{sec:RareB}, constraints from the rare decay $B \to X_s \gamma$ push the masses of the additional Higgs bosons to uninterestingly large values in the type~IIB and flipped~B models. We therefore forgo an in-depth discussion of constructing the mass matrices and couplings in those types.

\section{Constraints from Rare B Decays}\label{sec:RareB}

As discussed in the previous section, in the type~IB and lepton-specific~B models the neutral Higgs bosons couple to down type quarks in a flavor diagonal way. 
Many constraints from FCNCs in the down quark sector are therefore automatically avoided. 
There is one important exception: the $b \to s \gamma$ decay. We find that 1-loop contributions from charged Higgs bosons can lead to sizable NP effects in the $b \to s \gamma$ transition.\footnote{We also checked 1-loop charged Higgs contributions to the $B_s \to \mu^+\mu^-$ decay and tree level charged Higgs contributions to the $B \to \tau \nu$ and $B \to D^{(*)} \tau \nu$ decays and found that they are negligible in regions of parameter space that are allowed by $b \to s \gamma$.} Both the SM prediction~\cite{Misiak:2015xwa} and the experimental measurements of the $B \to X_s \gamma$ rate~\cite{Amhis:2016xyh} have uncertainties of less than $10\%$ and are in good agreement with each other, resulting in strong constraints on non-standard effects.

The new physics effects induced by charged Higgs loops can be described by modifications of the Wilson coefficients $C_7$ and $C_8$ of an effective Hamiltonian
\begin{eqnarray}
 \mathcal H_\text{eff}^\text{NP} = -\frac{4 G_F}{\sqrt{2}} V_{tb} V_{ts}^* \frac{e^2}{16\pi^2} \left( \Delta C_7 Q_7 + \Delta C_8 Q_8 \right) ~,
\end{eqnarray}
with the dipole operators
\begin{equation}
 Q_7 = \frac{1}{e} m_b (\bar s \sigma_{\mu\nu} P_R b) F^{\mu \nu} ~, \quad Q_8 = \frac{g_s}{e^2} (\bar s \sigma_{\mu\nu} T^a P_R b) G_a^{\mu \nu} ~,
\end{equation}
Using the results from~\cite{Bobeth:1999ww} we find for the charged Higgs contribution in the type~IB and lepton-specific~B scenarios
\begin{subequations}
\begin{eqnarray}
\Delta C_7 &=& \frac{m_t^2}{m_{H^\pm}^2} f_7\!\left(\frac{m_t^2}{m_{H^\pm}^2}\right)~, \\
\Delta C_8 &=& \frac{m_t^2}{m_{H^\pm}^2} f_8\!\left(\frac{m_t^2}{m_{H^\pm}^2}\right)~.
\end{eqnarray}
\end{subequations}
In the type~IIB and flipped~B scenarios the Wilson coefficients are only determined up to model dependent $O(1)$ factors
\begin{subequations}
\begin{eqnarray}
\Delta C_7 &=& O(1) \times \left( \frac{m_t^2}{m_{H^\pm}^2} g_7\!\left(\frac{m_t^2}{m_{H^\pm}^2}\right) + \tan^2\beta \frac{m_t^2}{m_{H^\pm}^2} h_7\!\left(\frac{m_t^2}{m_{H^\pm}^2}\right) \right) ~, \\
\Delta C_8 &=& O(1) \times \left( \frac{m_t^2}{m_{H^\pm}^2}g_8\!\left(\frac{m_t^2}{m_{H^\pm}^2}\right) + \tan^2\beta \frac{m_t^2}{m_{H^\pm}^2} h_8\!\left(\frac{m_t^2}{m_{H^\pm}^2}\right) \right) ~.
\end{eqnarray}
\end{subequations}
The loop functions $f_{7,8}$, $g_{7,8}$, and $h_{7,8}$ are given in appendix~\ref{loops}.
Note that in our type~IB and lepton-specific~B scenarios the contributions are independent of $\tan\beta$, while the contributions in the type~IIB and flipped~B scenarios contain terms that are proportional to $\tan^2\beta$ and can become extremely large in regions of parameter space with large $\tan\beta$.
This is in contrast to 2HDMs with natural flavor conservation, where the contributions are proportional to $1/\tan^2\beta$ (type~I and lepton-specific) and independent of $\tan\beta$ (type~II and flipped). 

\begin{figure}[tb]
\begin{center}
\includegraphics[width=0.6\textwidth]{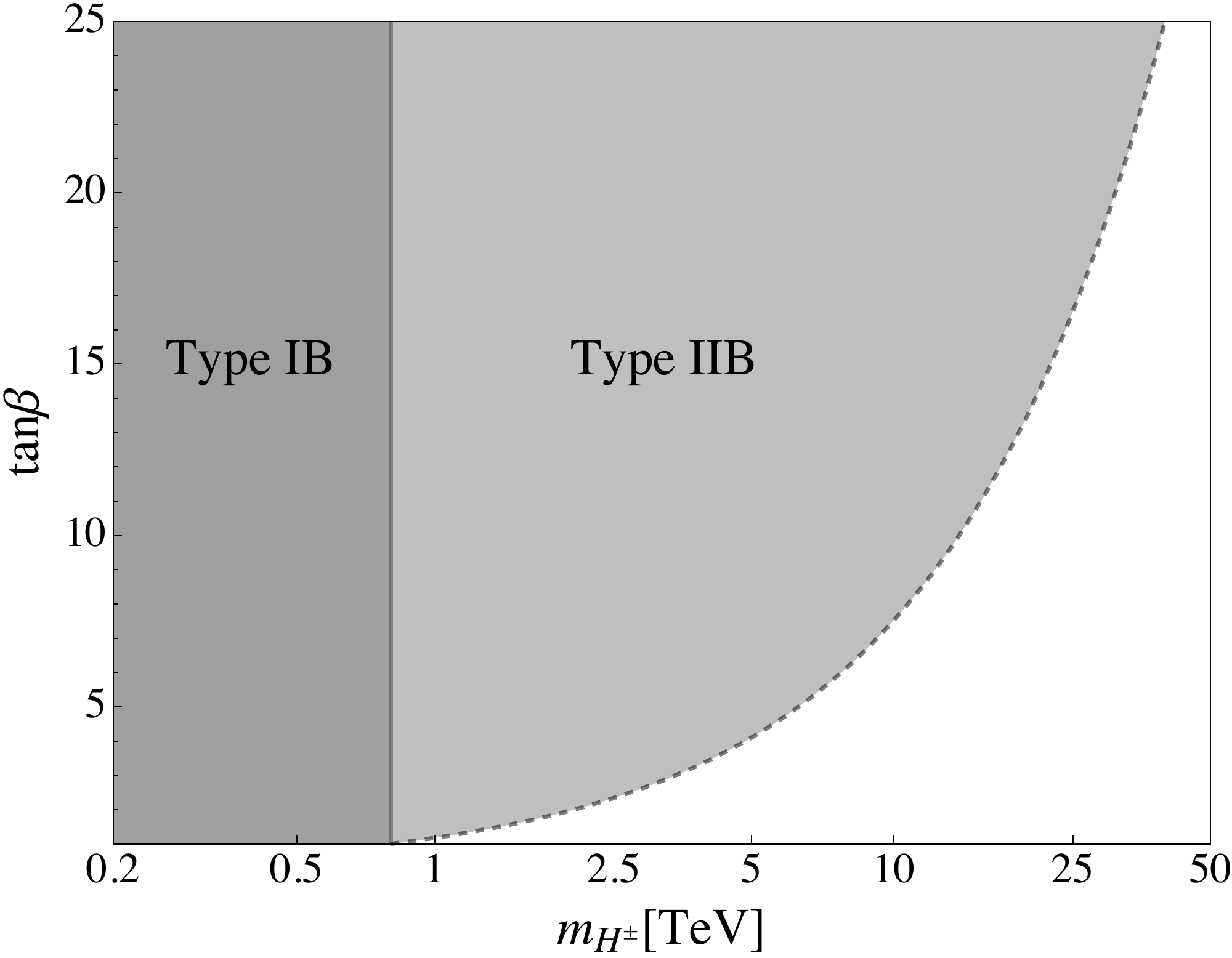}
\caption{Constraints from the $b \to s \gamma$ transition in the charged Higgs mass $m_{H^\pm}$ vs. $\tan\beta$ plane. The dark gray region is excluded in the type~IB and lepton-specific~B scenarios at the $95\%$ C.L. The light gray region is excluded in the type~IIB and flipped~B scenarios at the $95\%$ C.L.}
\label{fig:bsgamma}
\end{center}
\end{figure}

Using the constraints on the Wilson coefficients from $b \to s \gamma$ transitions derived in~\cite{Paul:2016urs} and taking into account 1-loop renormalization group running between the electroweak scale and the $b$ scale, we find at the $95\%$ C.L.
\begin{equation}
-0.032 < \eta^\frac{16}{23} \Delta C_7 + \frac{8}{3} \left(\eta^\frac{14}{23}-\eta^\frac{16}{23} \right) \Delta C_8 < 0.027 ~,
\end{equation}
with $\eta = \alpha_s(m_t)/\alpha_s(m_b) \simeq 0.52$. 

The corresponding constraints are shown in the plots of Fig.~\ref{fig:bsgamma} in the charged Higgs mass $m_{H^\pm}$ vs. $\tan\beta$ plane. In the case of the type~IB and lepton-specific~B models, we find a $\tan\beta$ independent bound on the charged Higgs mass of $m_{H^\pm} \gtrsim 800$~GeV. In the type~IIB and flipped~B models, we show as illustration the case where the free $O(1)$ parameters are set to exactly 1. In these types of models, the $b \to s\gamma$ constraint is highly dependent on $\tan\beta$, e.g. $m_{H^\pm} \gtrsim 800$~GeV for $\tan\beta =1$, but $m_{H^\pm} \gtrsim 15$~TeV for $\tan\beta =10$. Varying the $O(1)$ coefficients shifts the exclusion line up or down by an order one factor.

Note that because of the $SU(2)_L$ gauge symmetry, the masses of the heavy scalar and pseudoscalar Higgs differ from the charged Higgs mass only by a small amount: $m_H \simeq m_A \simeq m_{H^\pm}$ with splittings of the order of $v^2/m^2_{H^\pm} \lesssim 10\%$.
The bounds on the charged Higgs mass from $b \to s \gamma$ therefore hold approximately for the masses of the heavy scalar and pseudoscalar Higgs as well.

As discussed in Sec.~\ref{sec:F2HDM}, for the purpose of generating the fermion mass hierarchy we have in mind values of $\tan{\beta} \sim 1/\lambda_c \sim 5$ or $\tan{\beta} \sim 1/\lambda_c^2 \sim 25$. In the type~IIB and flipped~B models, we see that the $b \to s \gamma$ constraints therefore strongly disfavor Higgs bosons with masses at the TeV scale. 
This remains true even if we take into account generous choices of the free $O(1)$ parameters.
With this in mind we focus our remaining analysis on the type~IB and lepton-specific~B models.

\section{Rare Top Decays}\label{sec:RareTop}

In the SM, flavor-changing top quark decays $t \rightarrow hq$ are both loop and GIM suppressed, and are predicted to have very small branching ratios.
Using the results from~\cite{Eilam:1989zm,Abbas:2015cua} for the partial widths $\Gamma(t \rightarrow hq)$ and normalizing to the $t \to W b$ decay width which dominates the total top width, we derive the following compact expression for the branching ratios
\begin{equation}\label{eq:topBRSM}
 \text{BR}(t \to h q)_\text{SM} = \frac{G_F^2 m_b^4}{4 \pi^4} |V_{qb}|^2 \frac{\big(1-m^2_h/m^2_{t,\,\text{pole}}\big)^2}{\big(1-m^2_W/m^2_{t,\,\text{pole}}\big)^2 \big(1+2m^2_W/m^2_{t,\,\text{pole}}\big)} \mathcal F\left(\frac{m_t^2}{m_W^2},\frac{m_h^2}{m_W^2}\right) ~.
\end{equation}
The branching ratio is suppressed by four powers of the bottom mass, as expected from GIM. We use the bottom $\overline{\text{MS}}$ mass at the scale of the top $m_b(m_t) = 2.73$\,GeV.
Note that we are using the top pole mass in the phase space factors, but the top $\overline{\text{MS}}$ mass in the loop function $\mathcal F$. 
The explicit expression for $\mathcal F$ is given in the appendix~\ref{app:F}. For central values of the Higgs mass $m_h = 125.18$\,GeV~\cite{Tanabashi:2018oca} and the top $\overline{\text{MS}}$ mass $m_t(m_t) = 163.4$\,GeV (corresponding to a top pole mass of $m_{t,\,\text{pole}} = 173.0$\,GeV~\cite{Tanabashi:2018oca}) we find $\mathcal F \simeq 0.48$. 
By far the largest uncertainties in the rare top branching ratios are due to the CKM factors $|V_{cb}| = (42.2 \pm 0.8)\times 10^{-3}$~\cite{Tanabashi:2018oca} and $|V_{ub}| = (3.94 \pm 0.36)\times 10^{-3}$~\cite{Tanabashi:2018oca} and due to higher order QCD effects that we estimate by varying the renormalization scale of the bottom $\overline{\text{MS}}$ mass $m_b(\mu)$ in the range $m_t / 2 < \mu < 2 m_t$. We obtain %
\begin{subequations}
\begin{align}
\text{BR}(t\rightarrow hu)_\text{SM} &= \big(3.66^{+0.94}_{-0.70} \pm 0.67 \big) \times 10^{-17}\,, \\
\text{BR}(t \rightarrow hc)_\text{SM} &=  \big(4.19^{+1.08}_{-0.80} \pm 0.16 \big) \times 10^{-15}\,,
\end{align}
\end{subequations}
where the first uncertainty is due to the variation of the renormalization scale and the second is due to the CKM matrix elements. 
 The current strongest experimental bounds on these decays are obtained by the ATLAS experiment, using an integrated luminosity of 36 fb$^{-1}$ of $pp$ collision data with $\sqrt{s} = 13$ TeV in multi-lepton final state searches~\cite{Aaboud:2018oqm}, and read
\begin{subequations}
\begin{align}
\text{BR}(t\rightarrow hu) &< 0.12 \%  \\
\text{BR}(t \rightarrow hc) &< 0.11 \%,
\end{align}
\end{subequations}
The predicted values for these processes in the SM are far below the current sensitivities shown above. The projected sensitivities for the rare top decays at the high luminosity LHC (HL-LHC) for an integrated luminosity of 3 ab$^{-1}$ at 14 TeV are $\mathcal{O}(10^{-4})$~\cite{ATLAS:url,Cerri:2018ypt}. The projections for the Future Circular Collider (FCC) indicate sensitivities comparable to the HL-LHC for the $t \to hu$ decay and about an order of magnitude stronger for the $t \to hc$ decay~\cite{Mangano:2018mur}. The Compact Linear Collider (CLIC) could also place a limit comparable to the HL-LHC for the $t \to hc$ decay~\cite{deBlas:2018mhx}. 

The Yukawa textures in Sec.~\ref{sec:F2HDM} generate flavor-changing couplings for the SM-like Higgs boson, allowing for the rare top decays to appear at tree-level. Approximating the total width of the top quark by its dominant partial decay width to a $W$ boson and a $b$ quark, the branching ratios of the rare decays can be written as
\begin{align}\label{eq:topBR}
\text{BR}(t \rightarrow hq) &\simeq 2 |V_{qb}|^2 \frac{\cos^2(\beta - \alpha)}{\sin^2\beta \cos^2\beta}  \frac{\big(1-m^2_h/m^2_{t,\,\text{pole}}\big)^2}{\big(1-m^2_W/m^2_{t,\,\text{pole}}\big)^2 \big(1+2m^2_W/m^2_{t,\,\text{pole}}\big)} \nonumber \\
& \simeq \frac{\cos^2(\beta - \alpha)}{\sin^2\beta \cos^2\beta}  \times \begin{cases} 9.2 \times 10^{-4} ~~\text{for}~~ t \to h c ~, \\ 8.0 \times 10^{-6} ~~\text{for}~~ t \to h u ~. \end{cases}
\end{align}
As long as $\cos(\beta-\alpha) \neq 0$, the rare top decay branching ratios can be many orders of magnitude larger than the SM values, making these processes in our model accessible to current and future colliders. If $\cos(\beta-\alpha) = 0$ (the so called alignment limit) the couplings of the 125\,GeV Higgs are exactly SM-like. Deviations of $\cos(\beta-\alpha)$ from $0$ are constrained by measurements of Higgs production and decays at the LHC.  The constraints depend strongly on $\tan\beta$. In the appendix~\ref{app:fit} we show the allowed regions in the $\cos(\beta-\alpha)$ vs. $\tan\beta$ plane, taking into account all relevant LHC results on Higgs signal strength measurements.

\begin{figure}[tb]
\begin{center}
\includegraphics[width=0.46\textwidth]{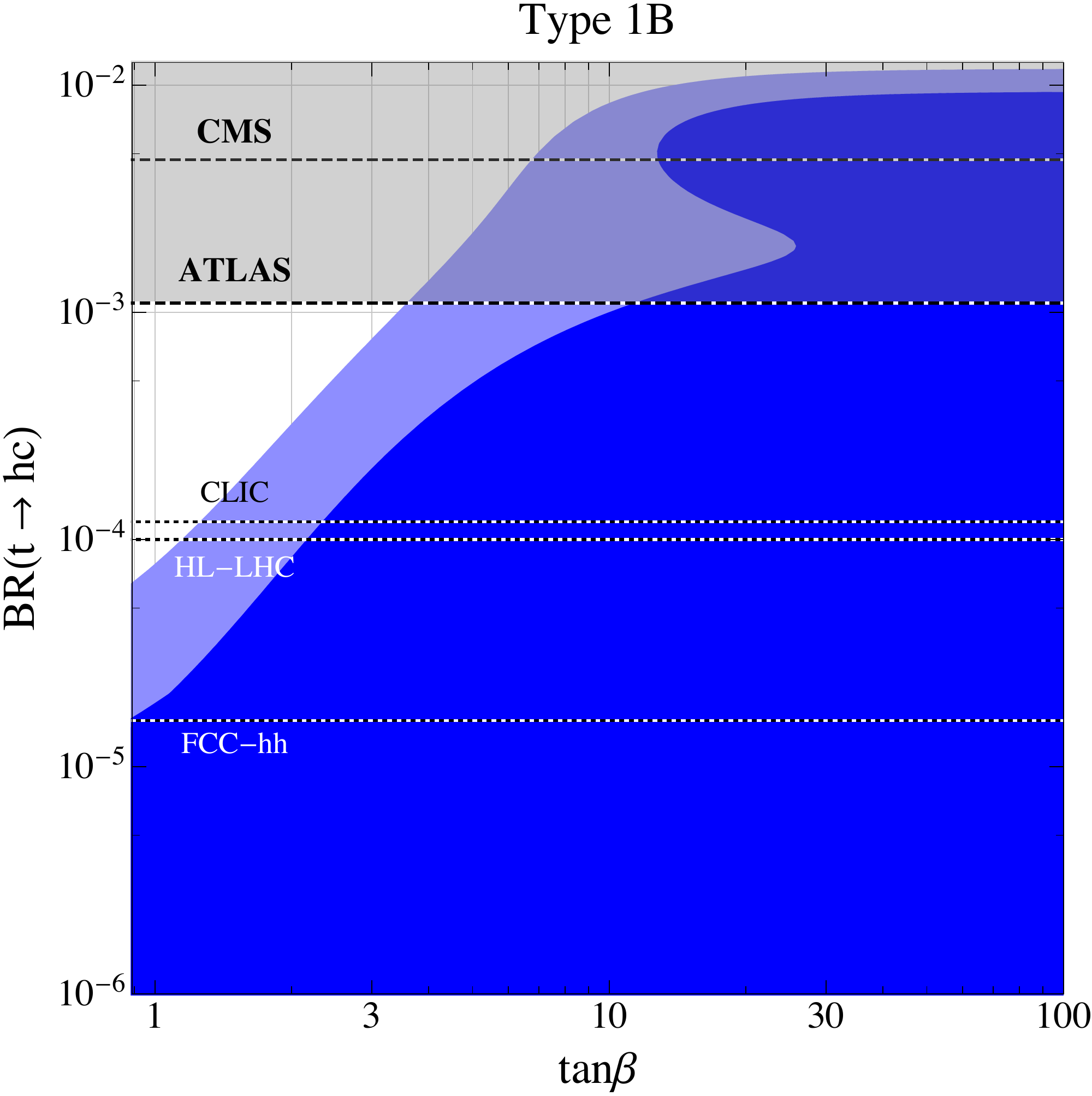}~~~~~~
\includegraphics[width=0.46\textwidth]{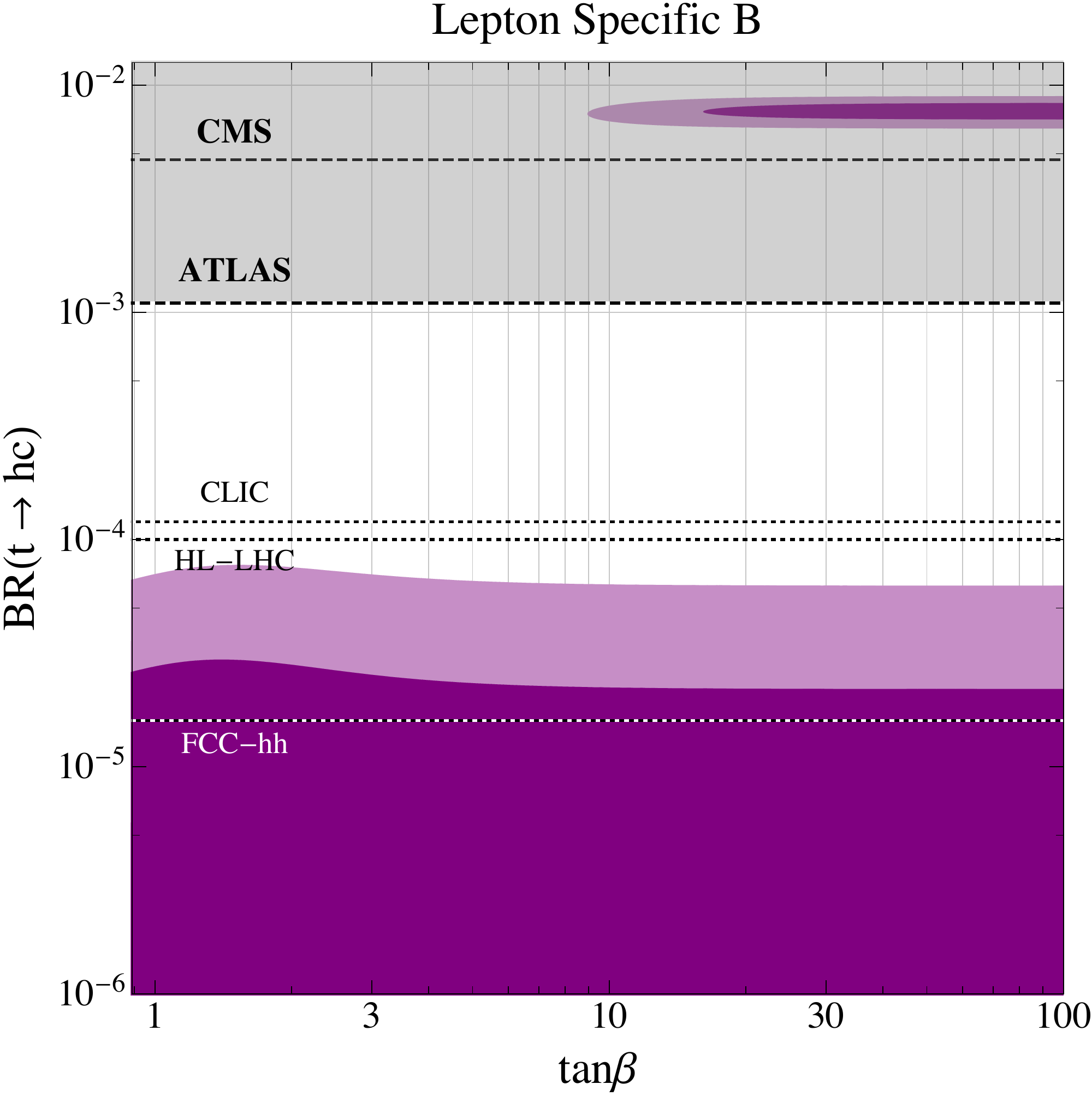} \\[12pt]
\includegraphics[width=0.46\textwidth]{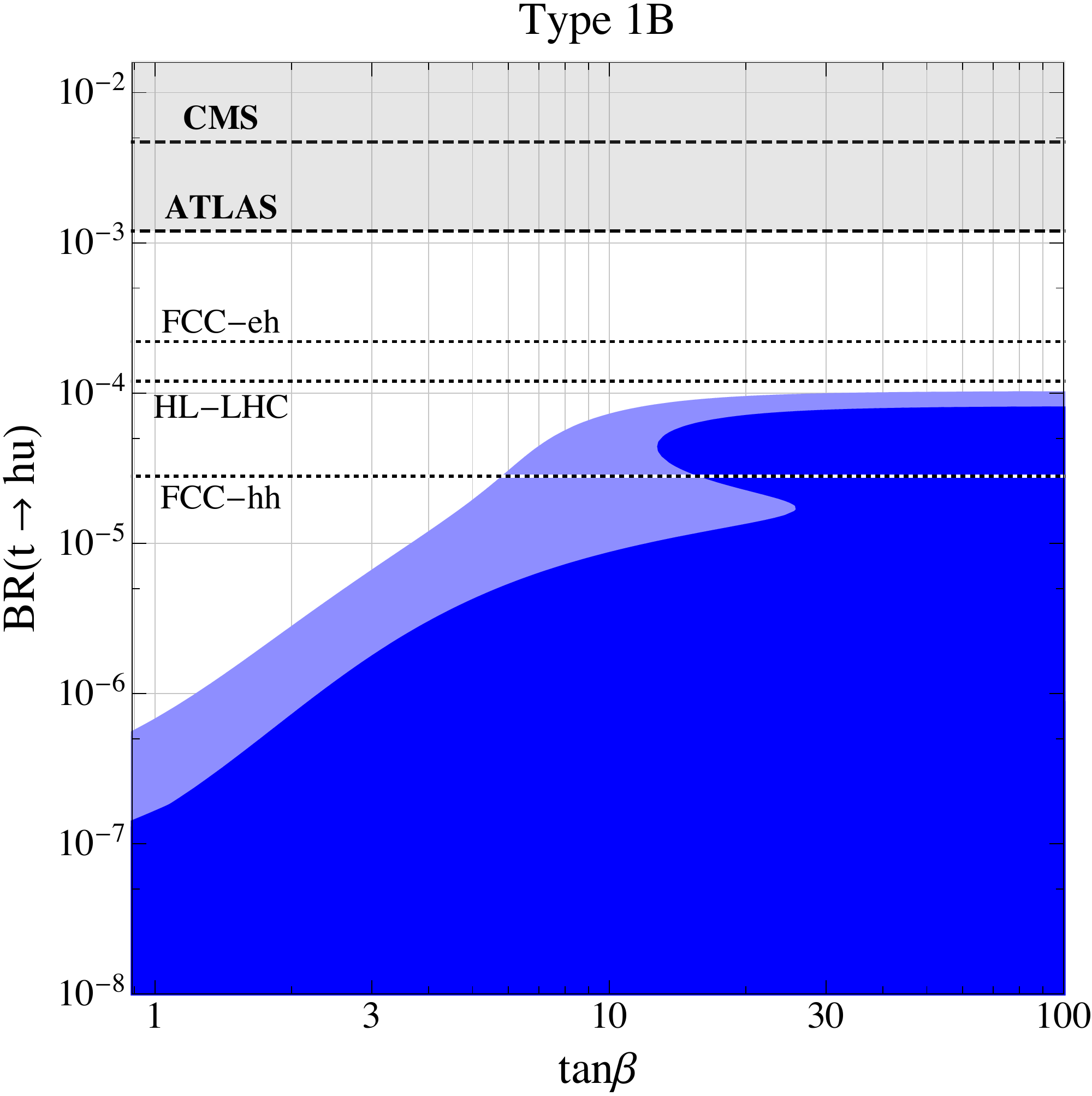}~~~~~~
\includegraphics[width=0.46\textwidth]{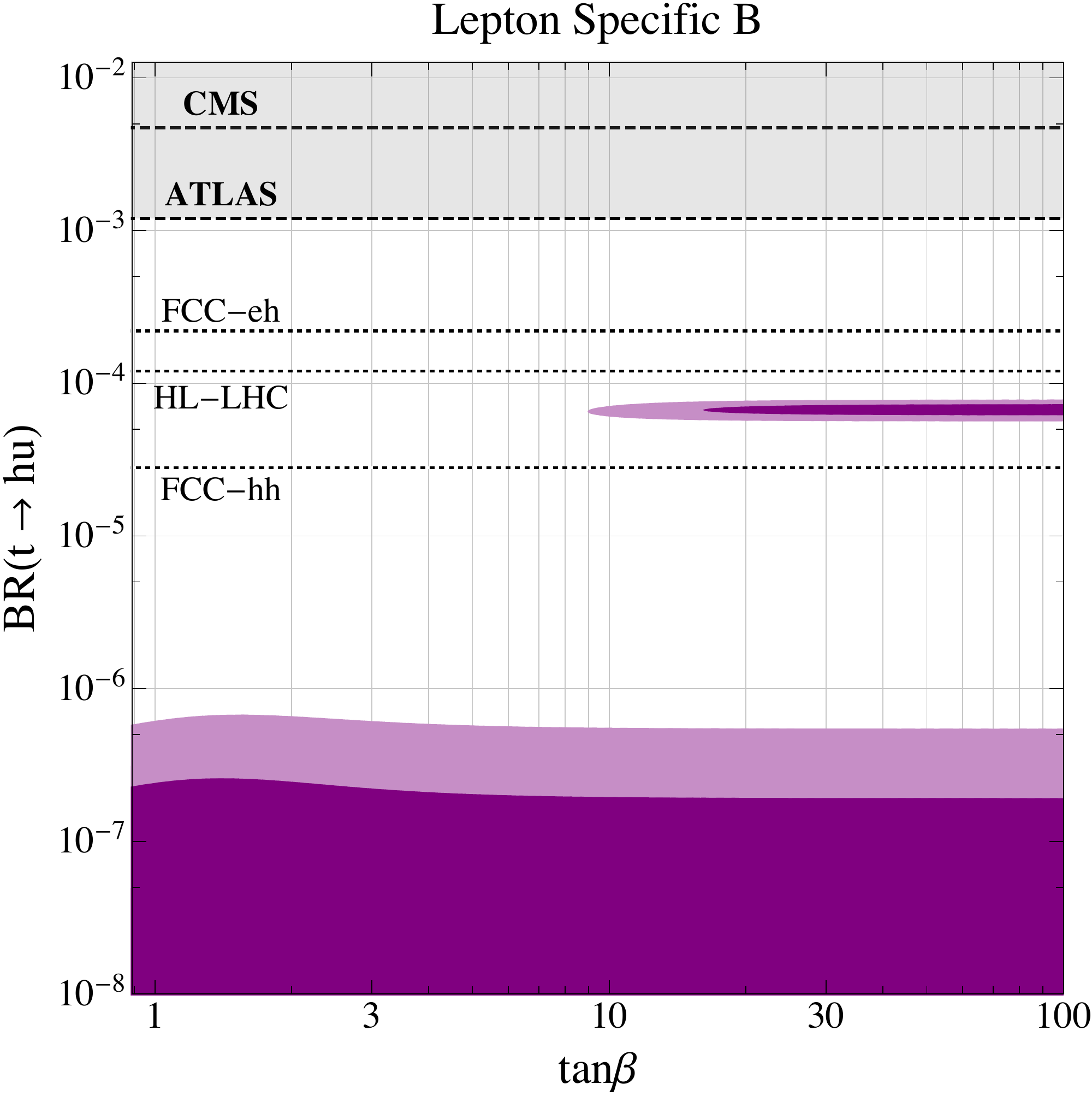}
\caption{The branching ratios $t \rightarrow hc$ (top) and $t \rightarrow hu$ (bottom) as a function of $\tan\beta$ 
in the type~IB model (left) and lepton-specific~B model (right). The blue and purple shaded regions are consistent with Higgs signal strength measurements. The dashed horizontal lines labeled ``ATLAS'' are the current best upper bounds on the branching ratios~\cite{Aaboud:2018oqm}. The dotted horizontal lines are the future projections from the HL-LHC, the FCC, and CLIC.}
\label{fig:topBR}
\end{center}
\end{figure}

In Fig.~\ref{fig:topBR} we use these allowed regions to give predictions for the rare top branching ratios as a function of $\tan\beta$ in the type~IB model (left) and lepton-specific~B model (right). The region in gray is excluded by the current ATLAS limits, while the dotted horizontal lines correspond to projected sensitives from the HL-LHC~\cite{Cerri:2018ypt}, the FCC~\cite{Mangano:2018mur}, and CLIC~\cite{deBlas:2018mhx}.

In the lepton-specific~B model, we observe two disjoint regions of parameter space. The upper region opens up for $\tan\beta \gtrsim 10$ and corresponds to a scenario where some of the Higgs couplings differ from the SM prediction by a sign, but are otherwise equal in magnitude. Such a scenario predicts BR($t\rightarrow hu) \simeq 6 \times 10^{-5}$ and BR($t\rightarrow hc) \simeq 7 \times 10^{-3}$ and is already excluded by the existing LHC constraints from~\cite{Aaboud:2018oqm}.

In general our models can give values for BR($t\rightarrow hu$) and BR($t\rightarrow hc$) that are much larger than the SM prediction, and can be in reach of current or future experimental sensitivities. 
In the case of $t \rightarrow hu$, the current LHC constraint from~\cite{Aaboud:2018oqm} does not probe the parameter space of our model. Also future projections from the the HL-LHC are unlikely to probe our model as they barely touch the region of predicted branching ratio values. The FCC-hh will start to cut into interesting parameter space of $t\rightarrow hu$ with a projected sensitivity of the order $10^{-5}$.

For the $t \rightarrow hc$ decay channel, the current LHC constraints already probe part of our model parameter space for moderate to large values of $\tan\beta \gtrsim 10$. Projections from the HL-LHC and CLIC will be also sensitive to parameter space with much lower choices of $\tan{\beta}$.

\section{Related Signatures}\label{sec:signatures}

Although the primary motivation for this model is to explore enhanced $t\to hq$ decays, the flavor structure of the up Yukawa couplings leads to other interesting features and signatures. We examine possible effects on $D$ meson mixing that arise from tree-level exchange of neutral Higgs bosons. We also consider the collider phenomenology of the heavy neutral and charged Higgs bosons ($H$, $A$, $H^\pm$), identifying the most prominent production and decay modes.

\subsection{Enhanced D Meson Mixing from Flavorful Higgs Bosons}\label{sec:MM}

In the SM, $D^0 - \bar{D}^0$ mixing proceeds through loop diagrams and is parameterized by the absolute values of the dispersive and absorptive part of the mixing amplitude, $x_{12} = 2 |M_{12}^D| \tau_D$, $y_{12} = |\Gamma_{12}^D| \tau_D$, and their relative phase $\phi_{12} = \text{Arg}(M_{12}^D/\Gamma_{12}^D)$, where $\tau_D$ is the lifetime of the $D^0$ meson.

The current world averages for the mixing parameters are~\cite{Amhis:2016xyh}
\begin{equation}\label{eq:x12}
x^\text{exp}_{12} = (0.43^{+0.10}_{-0.11})\% ~,~~~ y^\text{exp}_{12} = (0.63 \pm 0.06)\% ~,~~~ \phi_{12}^\text{exp} = (-0.25^{+0.96}_{-0.99})^\circ ~.
\end{equation}
In our model, we generically predict tree level Higgs contributions to $D^0 - \bar{D}^0$ mixing.
The corresponding effect on the dispersive part of the mixing amplitude is given by
\begin{align}
M^D_{12} = m^3_D \frac{f^2_D}{v^2_\text{w}}\frac{(V_{cb}V_{ub}^*)^2}{s^2_\beta c^2_\beta} & \bigg[\frac{1}{4}B_4 \eta_4 \bigg( \frac{c^2_{\beta - \alpha}}{m^2_h} + \frac{s^2_{\beta -\alpha}}{m^2_H} + \frac{1}{m^2_A}\bigg) \frac{m_u}{m_c} \nonumber \\
& - \bigg(\frac{5}{24} B_2 \eta_2 - \frac{1}{48}B_3 \eta_3\bigg)\bigg( \frac{c^2_{\beta - \alpha}}{m^2_h} + \frac{s^2_{\beta -\alpha}}{m^2_H} - \frac{1}{m^2_A}\bigg) \bigg] ~,
\end{align}
where the decay constant of the $D^0$ meson is $f_D \simeq 212$\,MeV~\cite{Aoki:2019cca}, the bag parameters are $B_2 \simeq 0.65$, $B_3 \simeq 0.96$, $B_4 \simeq 0.91$~\cite{Carrasco:2015pra}, and the 1-loop renormalization group factors are $\eta_2 \simeq 0.68$, $\eta_3 \simeq -0.03$, $\eta_4 = 1$~\cite{Altmannshofer:2018bch}. The absorptive part $\Gamma_{12}$ is unaffected in the model.
The results for the mixing amplitude are independent of the type of F2HDM, they hold both in type~IB and in the lepton-specific~B model.

Despite the fact that the neutral Higgs bosons contribute to $D^0 - \bar{D}^0$ mixing at tree level, the approximate $SU(2)^5$ flavor symmetry of the F2HDMs ensures that their contributions are very small, suppressed by small quark masses and CKM matrix elements.
For Higgs boson masses around 1\,TeV and values of $\tan\beta$ as large as 100, we find that the NP contribution to $D^0 - \bar{D^0}$ mixing is much smaller than the uncertainties in Eq.~\ref{eq:x12}. Improvements in precision by more than two orders of magnitude would be required to become sensitive to the predicted non-standard effects in our models.

\subsection{Collider Phenomenology of Heavy Higgs Bosons}\label{sec:Pheno}
 
The F2HDMs considered here offer a rich set of phenomenological consequences. Not only do these models predict additional Higgs bosons that could be within reach of the LHC but the introduction of tree-level FCNCs means that we anticipate distinct signatures coming from the new Higgs bosons that set this model apart from more traditional 2HDMs. 

\subsubsection{Heavy Higgs Production and Decays}
 
\begin{figure}[tb]
\begin{center}
\includegraphics[width=0.7\textwidth]{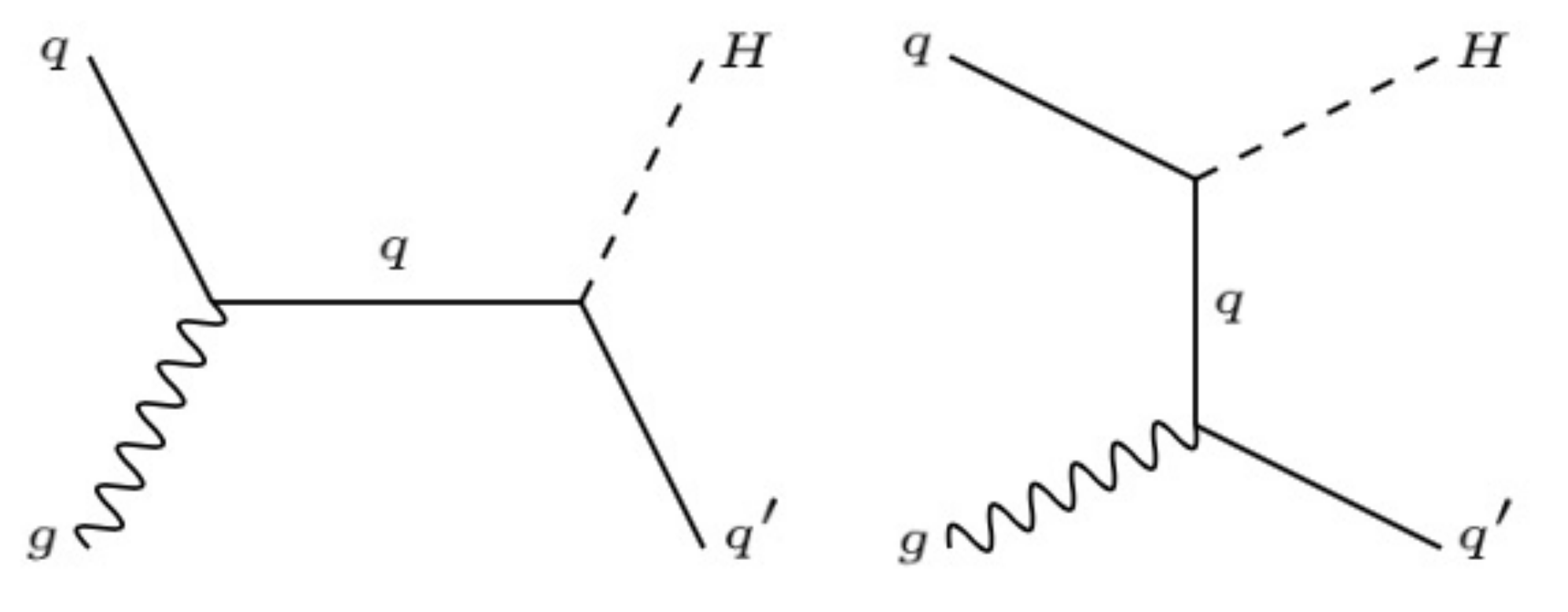} 
\caption{The Feynman diagrams for quark associated production of the heavy and charged Higgs bosons. In the context of F2HDMs this production mode can have a sizeable cross section due to the tree level flavor-changing neutral currents.}
\label{tHProdFeynman}
\end{center}
\end{figure}

There are several production modes via which the heavy Higgs bosons can be produced at the LHC. Due to the enhanced off-diagonal couplings in the up quark sector we expect top associated production, see Fig.~\ref{tHProdFeynman}, to contribute with a sizable cross section. 
 
\begin{figure}[tb]
\centering 
\includegraphics[width=0.475\textwidth]{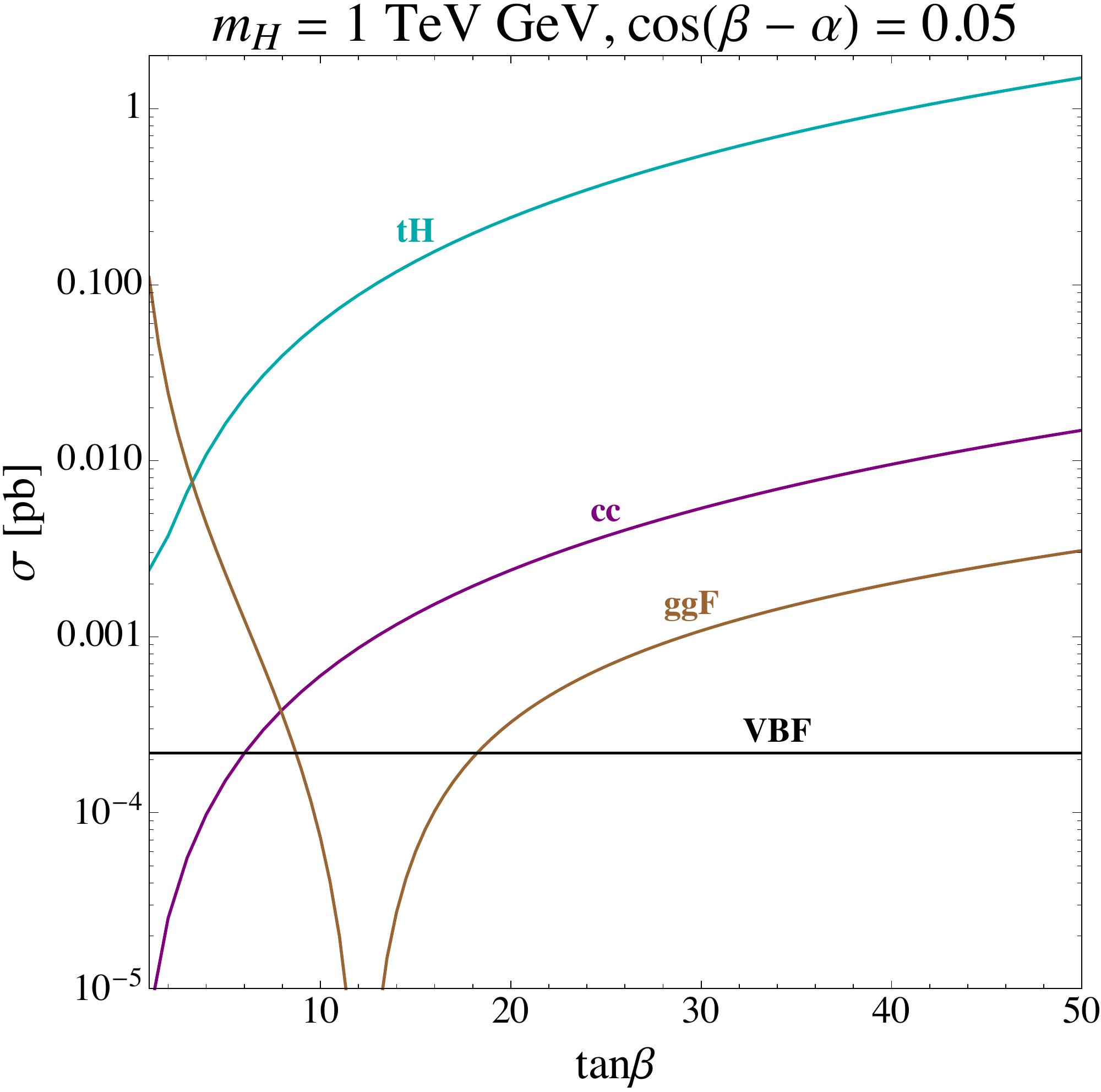}~~~~
\includegraphics[width=0.475\textwidth]{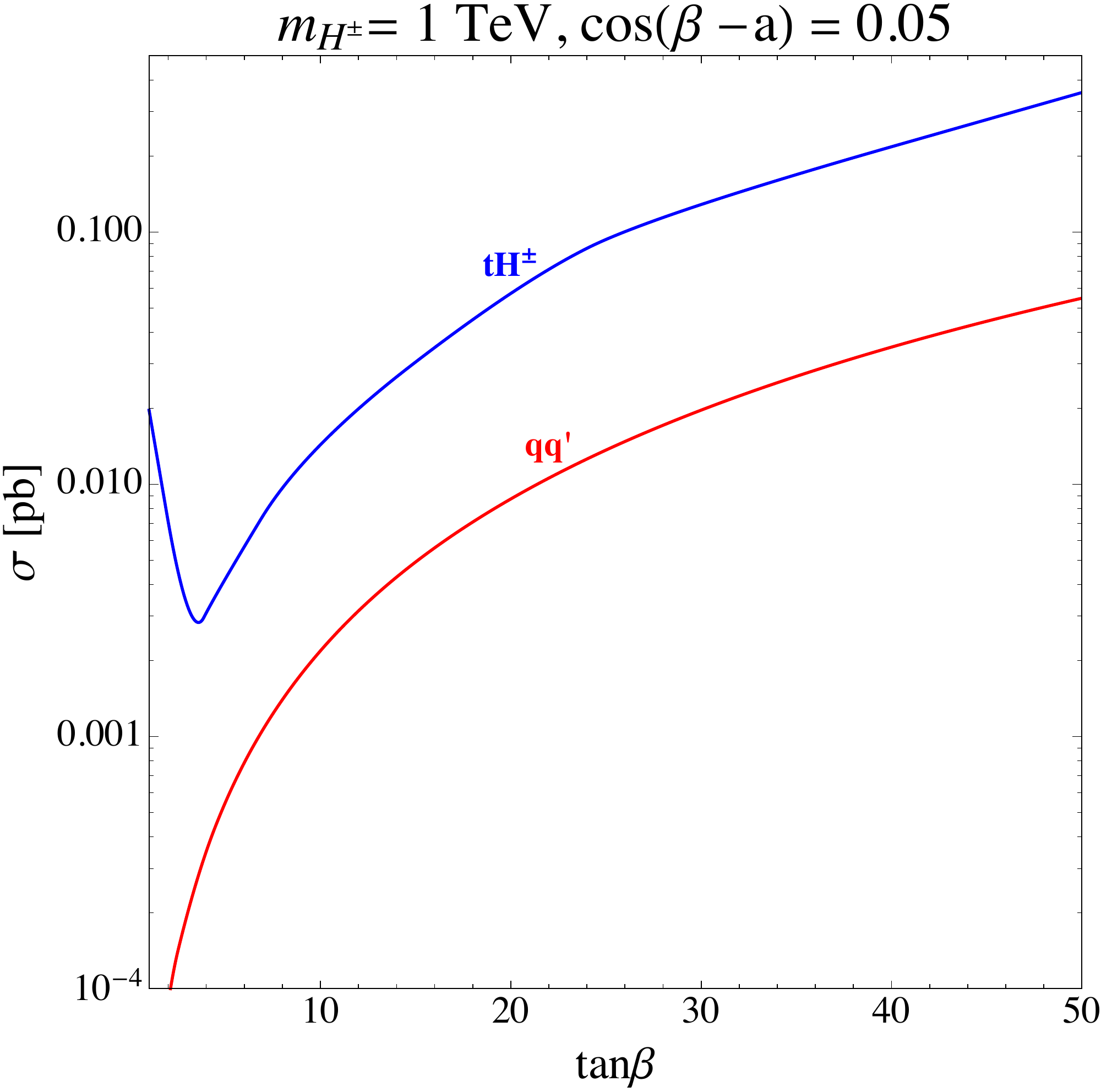}  \\[12pt] 
\includegraphics[width=0.475\textwidth]{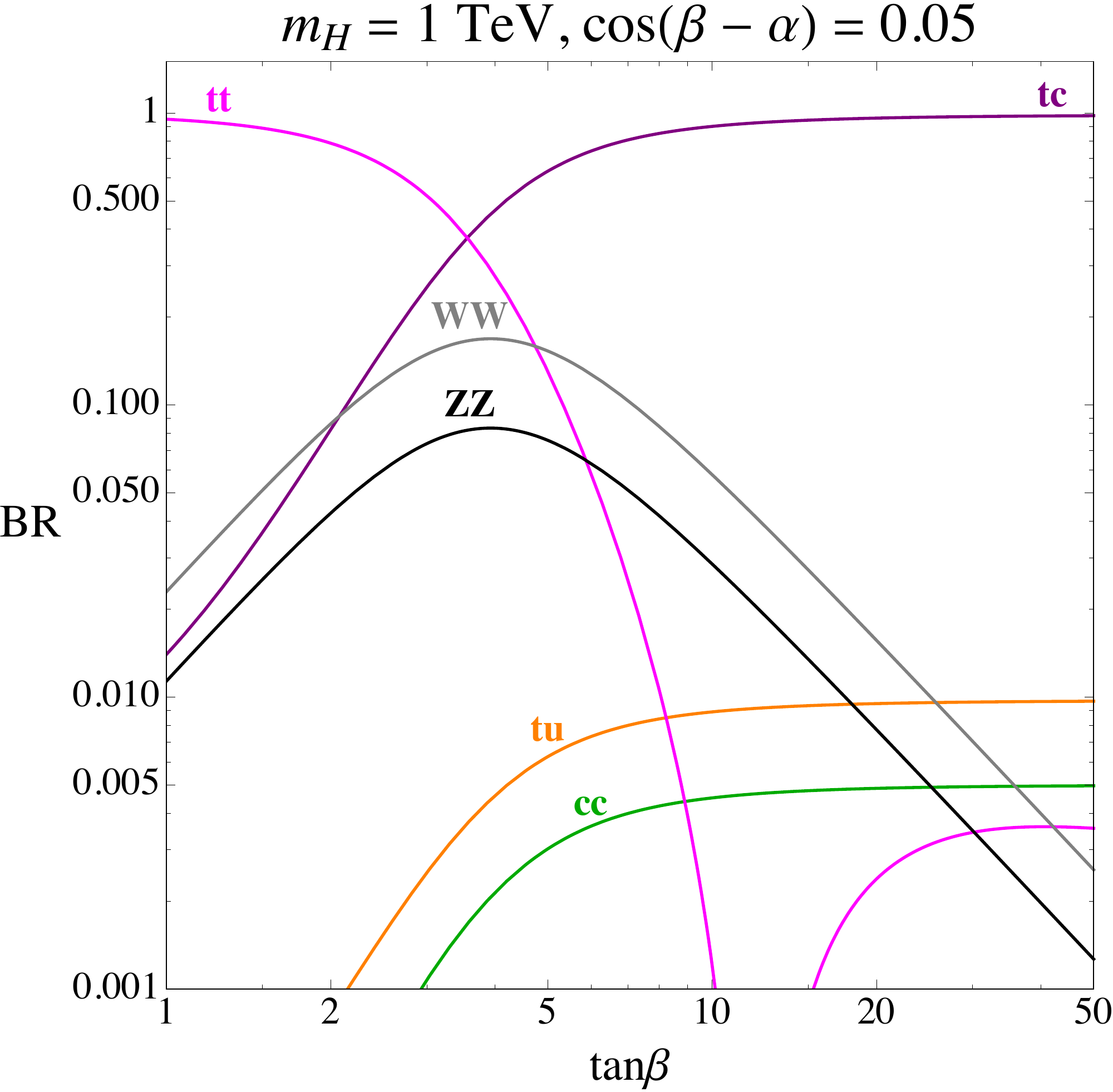} ~~~~
\includegraphics[width=0.475\textwidth]{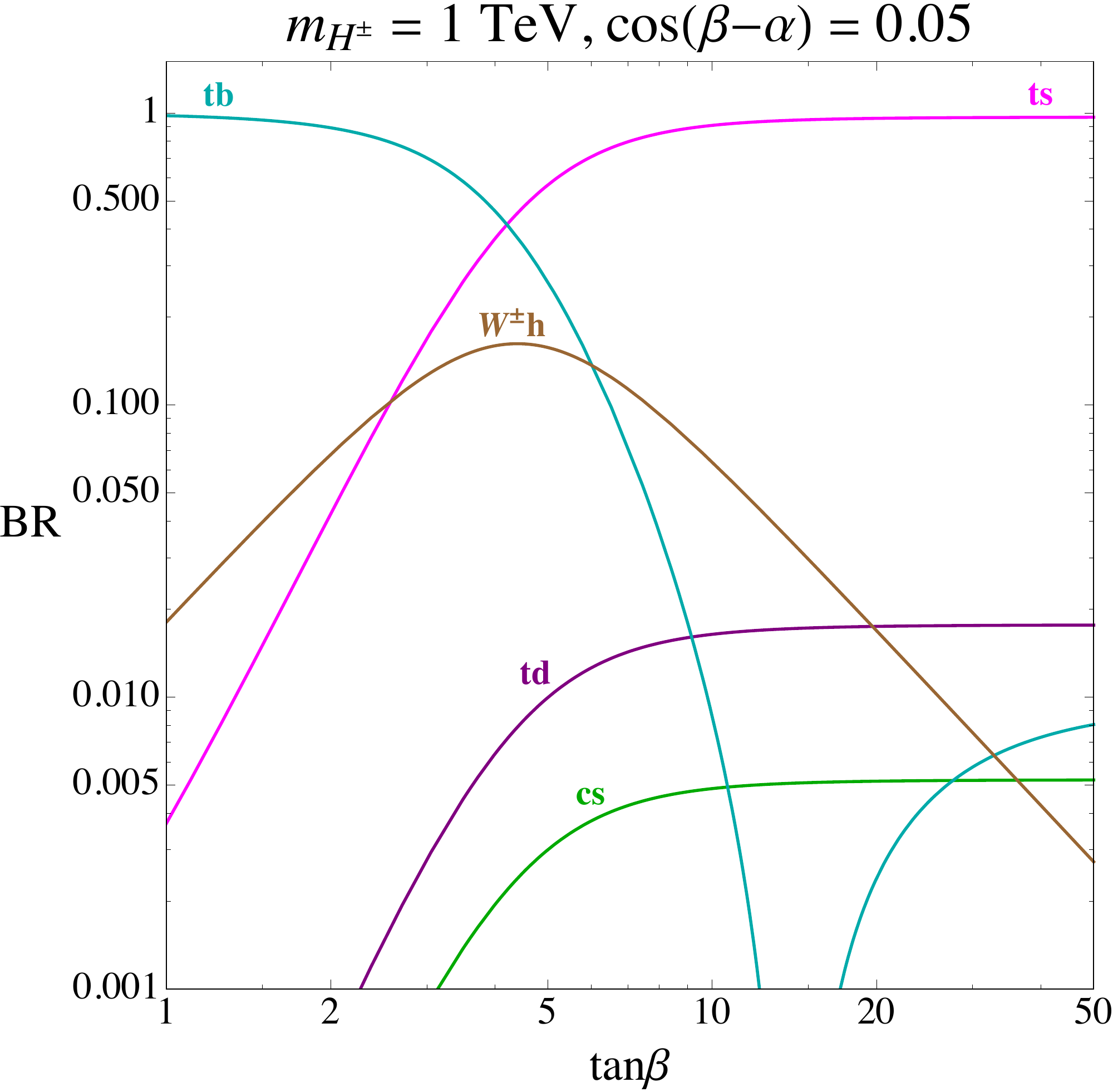}
\caption{Production cross sections (top) and branching ratios (bottom) of the heavy neutral Higgs $H$ (left) and the charged Higgs $H^\pm$ (right) in the flavorful 2HDM of type~IB as a function of $\tan\beta$ with the masses $m_H$ and $m_{H^\pm}$ fixed to 1~TeV and $\cos(\beta - \alpha) = 0.05$.}
\label{fig:sigmaBR}
\end{figure}

In order to evaluate the production cross sections, we follow the steps described in \cite{Altmannshofer:2016zrn,Altmannshofer:2018bch}. 
The results for the production cross sections are identical for the type~IB and lepton-specific~B models. 
The cross sections of the heavy neutral Higgs $H$ and the charged Higgs $H^\pm$ at 13~TeV proton-proton collisions as a function of $\tan\beta$ for fixed Higgs masses of 1\,TeV and $\cos(\beta - \alpha) = 0.05$ are shown in the upper plots of Fig.~\ref{fig:sigmaBR}. 
Small values of $\cos(\beta - \alpha) \ll 1$ are motivated by the constraints from Higgs signal strength measurements (see appendix~\ref{app:fit}).

For the neutral scalar $H$, associated production with a top and gluon-gluon fusion are the dominant production modes. At low $\tan{\beta}$ gluon-gluon fusion is largest because the coupling to tops is unsuppressed. As $\tan{\beta}$ increases the gluon-gluon fusion rate drops and is overtaken by top associated production which is enhanced for large $\tan{\beta}$. The dominant production cross sections for the heavy pseudoscalar $A$ are almost identical to those of the heavy Higgs.
For the charged Higgs, top associated production is the largest production mechanism over the full range of $\tan\beta$ values.

The branching ratios of the heavy neutral Higgs $H$ and the charged Higgs $H^\pm$ are shown in the lower plots Fig.~\ref{fig:sigmaBR}
in the type~IB model. Results in the lepton-specific~B model are almost identical. The main difference in the lepton-specific~B model is the presence of a $\tau\tau$ branching ratio at the level of few percent, which is strongly suppressed in the type~IB model.

As expected, for moderate to large $\tan{\beta}$ the dominant decay mode of $H$ is the flavor-changing $H \to ct$. The branching ratio to $t\bar t$ can be substantial, however this decay mode primarily plays a role for small $\tan{\beta}$ which is less motivated by the quark mass hierarchy. For moderate $\tan{\beta}$ we also notice that the gauge bosons can contribute at a level between $1-10\%$. 
The branching ratios of the charged Higgs tend to be dominated by $ts$ and $tb$ decays. In particular, for low $\tan{\beta}$ $tb$ dominates. Once $\tan{\beta}$ becomes larger than about 5 we see $ts$ dominates for the rest of the parameter space. In addition to the most dominant decays, we see that at the level of a few percent or lower we can expect decays to $W^{\pm} h$.

\subsubsection{Collider Signatures}

The constraints on this model from existing searches for heavy Higgs boson are very weak due to the unique flavor structure. The overwhelming decay of the neutral Higgses to $ct$ means the branching ratio to other modes is highly suppressed as seen in Fig.~\ref{fig:sigmaBR}. Typical search channels at the LHC are through these suppressed channels, such as $\mu\mu$, $\tau\tau$, $VV$, and $jj$, making the prospects of discovering a heavy Higgs through these channels very weak. Also, the standard searches for charged Higgs bosons in the $\tau\nu$ channel hardly constrain our parameter space, due to the strongly reduced branching ratios $H^\pm \to \tau \nu$. Unique signatures that are relevant to collider searches of our model are driven by the large non-standard decay modes $H,A \to tc$, and  $H^{\pm} \to ts$.

\begin{figure}[tb]
\begin{center} 
\includegraphics[width=0.45\textwidth]{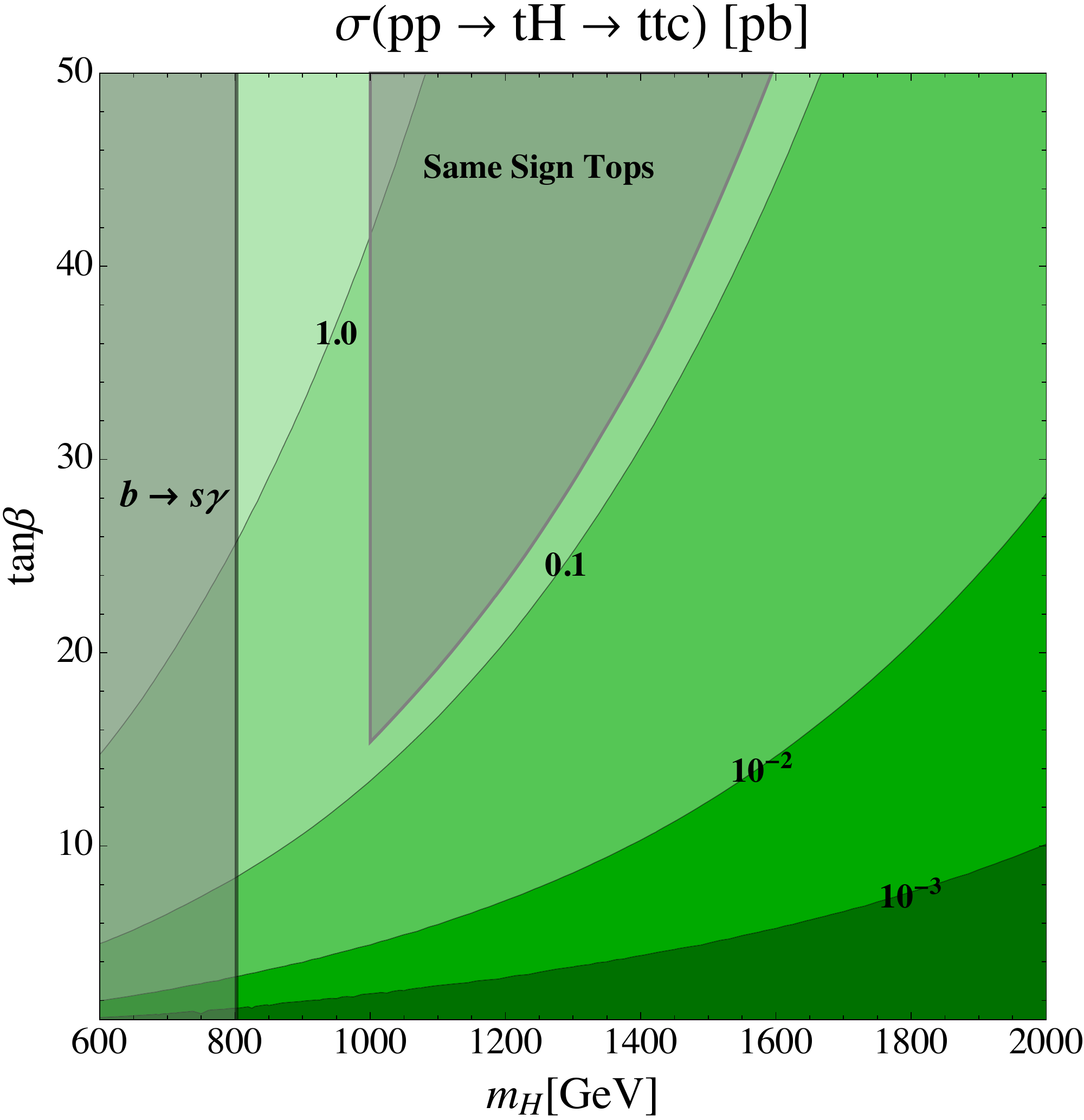} ~~~~~
\includegraphics[width=0.45\textwidth]{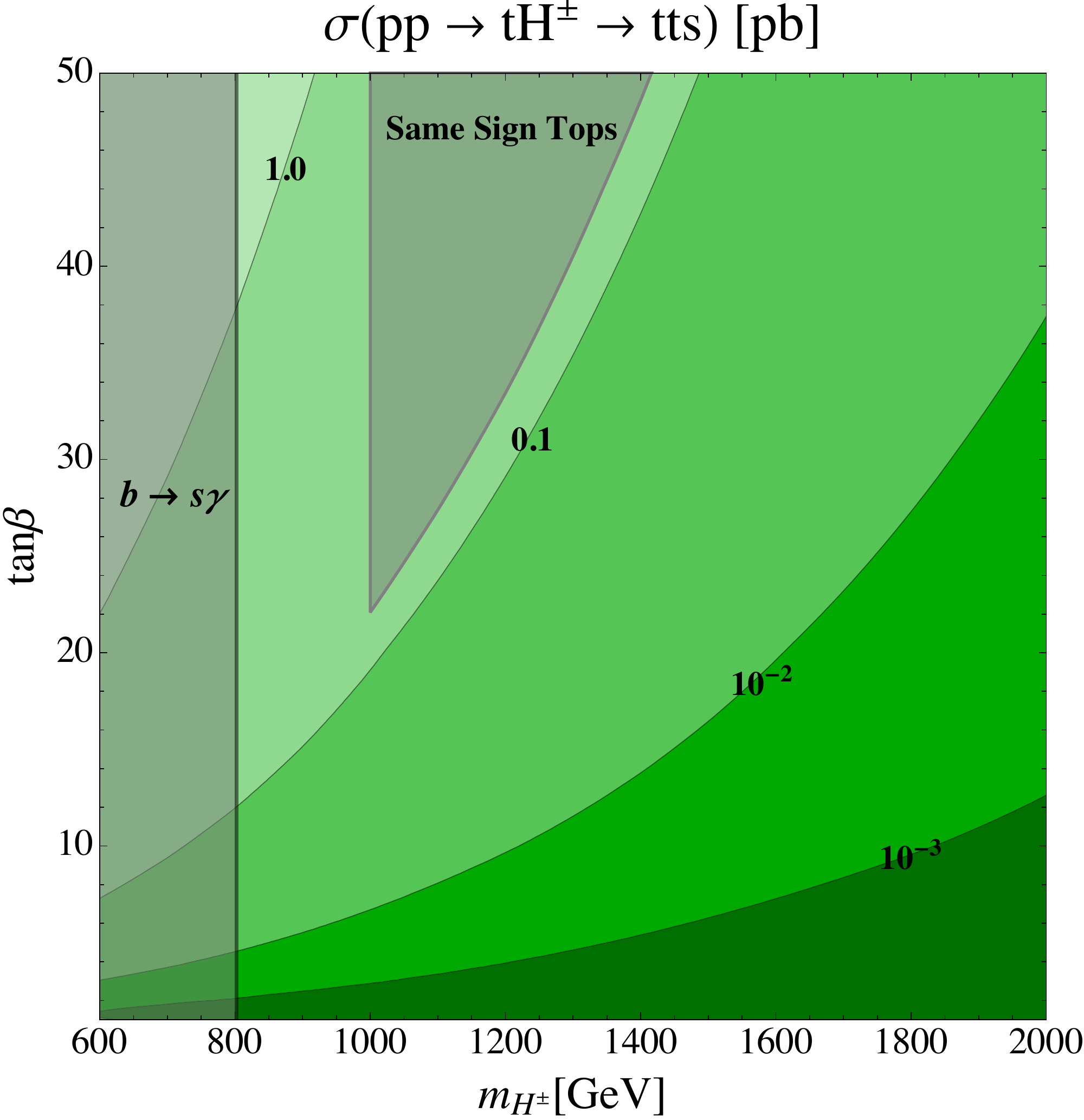} \\[16pt]
\caption{Cross section of same-sign tops plus jet from the production and decay of a neutral heavy Higgs $H$ (left) as well as opposite-sign tops plus jet from the production and decay of a charged Higgs (right) in the considered flavorful 2HDM of type~IB in the plane of Higgs mass vs. $\tan\beta$. The gray shaded regions for light Higgs masses are excluded by $b \to s \gamma$ constraints (see Sec.~\ref{sec:RareB}). The triangle shaped gray region for large $\tan\beta$ is excluded by existing searches for same-sign tops~\cite{Aaboud:2018xpj}. Throughout the plots we set $\cos(\beta-\alpha)  = 0.05$.}
\label{fig:tHtotc}
\end{center}
\end{figure}

The charged Higgs produced via top associated production and subsequent decay to $ts$ leads to opposite-sign tops that do not reconstruct a resonance. This is similar to charged Higgs bosons in 2HDMs with natural flavor conservation. The unique feature with respect to 2HDMs with natural flavor conservation is that the accompanying jet of the $t\bar t$ system is not a $b$-jet but a strange jet.
The cross section for the $t\bar t$ + jet signature as a function of $\tan{\beta}$ and charged Higgs mass is shown in the right plot of Fig.~\ref{fig:tHtotc}. We find cross sections that can easily exceed 100\,fb for Higgs masses of $O$(1\,TeV) and sizable $\tan\beta$.
The shaded region to the left of the vertical line at Higgs masses of around 800\,GeV is excluded by the constraint from $b \to s \gamma$. Existing searches for charged Higgs bosons that decay to $tb$~\cite{Aaboud:2018cwk,CMS:1900zym} make heavy use of b-tagging and are therefore not directly applicable in our scenario. Our work motivates dedicated studies of the $pp \to t H^- \to t \bar t s$ signature.

The heavy neutral Higgs being produced through top associated production along with a decay into $tc$ leads also to a final state with di-tops that do not reconstruct a resonance. In our flavorful 2HDMs, $50\%$ of the time the final state will be {\it same-sign} tops, in contrast to 2HDMs with natural flavor conservation that only produce opposite sign tops. 
Same-sign tops have been identified as important probes in a number of new physics scenarios, including RPV SUSY \cite{Durieux:2013uqa}, 2HDMs \cite{Kim:2015zla, Gori:2017tvg, Hou:2018zmg}, additional scalars \cite{BarShalom:2008fq, Chivukula:2013hga}, colored vectors \cite{Zhang:2010kr}, and effective field theories \cite{Atwood:2013xg}. The cross section of same-sign tops in our scenario is shown in the left plot of Fig.~\ref{fig:tHtotc} in the plane of Higgs mass vs. $\tan\beta$.
For Higgs masses of $O$(1\,TeV) we find cross sections up to 1\,pb. 
The shaded region to the left of the vertical line at Higgs masses of around 800\,GeV is excluded by the constraint from $b \to s \gamma$ assuming that $m_H \simeq m_{H^\pm}$.

In~\cite{Aaboud:2018xpj} searches for same-sign leptons are interpreted in a benchmark model in which same-sign tops are created by a neutral spin-1 mediator. Assuming that the acceptances and efficiencies are comparable in our scenario with a scalar mediator, we find that the large $\tan\beta$ region is already partly probed by the existing search. We show the region that is excluded by the same-sign top search also in the charged Higgs plot, assuming that $m_H \simeq m_{H^\pm}$. Keeping in mind that our $pp \to H t \to tt \bar c$ cross section approximately scales as $\tan^2\beta$, we expect that parameter space with $\tan\beta$ as low as $\sim 10$ might be probed by same-sign top searches at the high luminosity LHC.

\section{Conclusions}\label{sec:conclusion}

Rare top decays are strongly suppressed in the SM and their observation at existing or planned colliders would be a clear indication of new physics. One new physics framework that can lead to branching ratios of $t \to h c$ and $t\to h u$ in reach of current or future colliders are flavorful 2HDMs. 

In this work we explored a version of flavorful 2HDM where quark mixing dominantly resides in the up quark sector, leading to FCNCs in the up quark sector at tree level. We constructed a flavor model based on $U(1)$ flavor symmetries which successfully reproduces the measured quark masses and CKM mixing angles. We find that our model is highly predictive as the flavor structure of all Higgs couplings is fully determined by the quark masses and CKM matrix elements.

We give predictions for $t \to hc$ and $t \to hu$ rates in our model and show that the branching ratios can reach values of BR$(t \to hc) \sim 10^{-2}$ and BR$(t \to h u) \sim 10^{-4}$ (see Fig.~\ref{fig:topBR}) without violating constraints from Higgs signal strength measurements at the LHC. Existing bounds on BR$(t \to hc)$ from the LHC already start to constrain model parameter space. Expected sensitivities at the high-luminosity LHC or future colliders will be able to probe broad regions of parameter space.
In passing we also provide updated predictions for the $t\to hc$ and $t \to hu$ branching ratios in the SM (see Eq.~\ref{eq:topBRSM}). 

We explored additional effects of the up quark FCNCs in low energy flavor violating processes. In particular, we find that 1 loop effects in the rare $B$ decay $b \to s \gamma$ lead to strong constraints on the masses of the additional Higgs bosons of at least $\sim 800$\,GeV. On the other hand, constraints from $D$ meson mixing are weak in our setup. 

Finally, we explored the phenomenology of the heavy neutral and charged Higgs bosons of the F2HDM. We find that both neutral and charged Higgses are mainly produced in association with top quarks. The by far dominant decay modes are $tc$ and $ts$, respectively. These final states are not typical search channels of Higgs bosons in traditional 2HDMs. Therefore, current constraints from colliders are weak. 
The most prominent signatures of the models are $pp \to t H^- \to t \bar t s$, i.e. opposite sign tops + jet, and in particular $pp \to t H \to tt \bar c$, i.e. same-sign tops + jet. Cross sections of these signatures can be of the order of $100$\,fb for Higgs masses around 1\,TeV (see Fig.~\ref{fig:tHtotc}). Our results in the F2HDM framework motivate continued searches for same-sign tops and provides an additional benchmark model in which future same-sign top searches can be interpreted.

\section*{Acknowledgements}

We thank Stefania Gori for useful discussions.
The research of WA is supported by the National Science Foundation under Grant No. PHY-1912719.

\appendix
\section{Yukawa Couplings in the Quark Mass Eigenstate Basis}\label{mass_eigenstates}

In this appendix we show that in the considered type~IB and lepton-specific~B models, the couplings of the Higgs bosons to the quarks in the quark mass eigenstate basis are entirely determined by the known quark masses and CKM elements. 

The starting point are the Yukawa couplings $\lambda^q$ and $\lambda^{\prime q}$ in Eq.~\eqref{eq:yuk} that need to be rotated into the quark mass eigenstate basis. We perform unitary rotations on the left-handed and right-handed quark fields $q_{L/R} \to U_{q_{L/R}} q_{L/R}$ such that $U_{u_L}^\dagger (v \lambda^u + v^\prime \lambda^{\prime u}) U_{u_R} = \text{diag}(m_u,m_c,m_t) \equiv m^\text{diag}_u$ and analogous for the down quarks. Given the structure of $\lambda^u$ and $\lambda^{\prime u}$ in Eq.~\eqref{eq:lambda_u} we can introduce the matrix $\Pi = \text{diag}(0,0,1)$ that leaves $\lambda^u$ invariant and that annihilates $\lambda^{\prime u}$: $\Pi\, \lambda^u = \lambda^u$ and $\Pi\, \lambda^{\prime u} = 0$~.
Using this matrix, we can express $\lambda^u$ in the mass eigenstate basis directly in terms of quark masses and the CKM matrix
\begin{equation}
 U_{u_L}^\dagger \lambda^u U_{u_R} = U_{u_L}^\dagger \Pi (\lambda^u + \frac{v^\prime}{v} \lambda^{\prime u}) U_{u_R} = \frac{\sqrt{2}}{v} U_{u_L}^\dagger \Pi\, U_{u_L} m^\text{diag}_u = \frac{\sqrt{2}}{v} V_\text{CKM} \Pi\, V_\text{CKM}^\dagger m^\text{diag}_u~.
\end{equation}
In the last step we used the definition of the CKM matrix $V_\text{CKM} = U_{u_L}^\dagger U_{d_L}$ and the fact that $U_{d_L}$ and $\Pi$ commute due to the structure of $\lambda^d$ and $\lambda^{\prime d}$ in Eq.~\eqref{eq:lambda_d}.
Analogously, we can use the matrix $\Pi^\prime = \text{diag}(1,1,0)$ to get an expression for $\lambda^{\prime u}$ in the mass eigenstate basis
\begin{equation}
 U_{u_L}^\dagger \lambda^{\prime u} U_{u_R} = U_{u_L}^\dagger \Pi^\prime (\frac{v}{v^\prime} \lambda^u + \lambda^{\prime u}) U_{u_R} = \frac{\sqrt{2}}{v^\prime} U_{u_L}^\dagger \Pi^\prime\, U_{u_L} m^\text{diag}_u = \frac{\sqrt{2}}{v^\prime} V_\text{CKM} \Pi^\prime\, V_\text{CKM}^\dagger m^\text{diag}_u~.
\end{equation}
In the down quark sector we instead get 
\begin{eqnarray}
 && U_{d_L}^\dagger \lambda^d U_{d_R} = U_{d_L}^\dagger \Pi (\lambda^d + \frac{v^\prime}{v} \lambda^{\prime d}) U_{d_R} = \frac{\sqrt{2}}{v} U_{d_L}^\dagger \Pi\, U_{d_L} m^\text{diag}_d = \frac{\sqrt{2}}{v} \Pi\,   m^\text{diag}_d~, \\
 && U_{d_L}^\dagger \lambda^{\prime d} U_{d_R} = U_{d_L}^\dagger \Pi^\prime (\frac{v}{v^\prime} \lambda^d + \lambda^{\prime d}) U_{d_R} = \frac{\sqrt{2}1}{v^\prime} U_{d_L}^\dagger \Pi^\prime\, U_{d_L} m^\text{diag}_d = \frac{\sqrt{2}}{v^\prime} \Pi^\prime\, m^\text{diag}_d~.
\end{eqnarray}

\section{Loop Functions for $b \to s \gamma$}\label{loops}

In this appendix we give the explicit expressions for the loop functions that enter the results for the charged Higgs contributions to the $b \to s \gamma$ decay in section~\ref{sec:RareB}.

\begin{align}
 f_7(x) &= \frac{(2-3x)^2\log x}{12(1-x)^4} + \frac{11-43x+38x^2}{72(1-x)^3}~,  & \lim_{x\to0} f_7(x)= \frac{1}{3} \log(x) +\frac{11}{72}~, \\ 
 f_8(x) &= \frac{(2-3x)\log x}{4(1-x)^4} + \frac{16-29x+7x^2}{24(1-x)^3}~, & \lim_{x\to0} f_8(x) = \frac{1}{2}\log(x) +\frac{2}{3}~, \\ 
 g_7(x) &= -\frac{x(2-3x)\log x}{12(1-x)^4} - \frac{7-5x-8x^2}{72(1-x)^3}~, & \lim_{x\to0} g_7(x) = -\frac{7}{72}~, \\ 
 g_8(x) &= -\frac{x\log x}{4(1-x)^4} - \frac{2+5x-x^2}{24(1-x)^3}~, & \lim_{x\to0} g_8(x) = -\frac{1}{12}~,  \\ 
 h_7(x) &= \frac{(2-3x)\log x}{6(1-x)^3} + \frac{3-5x}{12(1-x)^2}~, & \lim_{x\to0} h_7(x) = \frac{1}{3}\log(x)+\frac{1}{4}~, \\ 
 h_8(x) &= \frac{\log x}{2(1-x)^3} + \frac{3-x}{4(1-x)^2}~, & \lim_{x\to0}h_8(x) = \frac{1}{2}\log(x)+ \frac{3}{4} ~.
\end{align}

\section{Loop Function for $t \to h q$}\label{app:F}

The loop function $\mathcal F$ that enters our SM expression of the rare top branching ratios Eq.~\eqref{eq:topBRSM} can be written as
\begin{align} \label{eq:F}
 \mathcal F(x,y) =& \frac{1}{16(x-y)^2}\Bigg|  x \Big(B_0(0,0,1)-B_0(x,0,1)\Big)+y \Big(B_0(y,0,0)-B_0(0,0,1) \Big) \\
 & 4\Big(B_0(x,0,1)-B_0(y,0,0)\Big)+(2+y)\Big(B_0(y,1,1)-B_0(x,1,0)\Big) \nonumber \\
 & -(4-2x+y)C_0(y,x,0,0,0,1)+(2-y+y^2-x(2+y))C_0(y,x,0,1,1,0)  \nonumber \\
 & +2(y-x)B^\prime_0(0,1,0)+2(x-2)B^\prime_0(x,1,0)+2(x^2+y-xy-2)C^\prime_0(y,x,0,1,1,0)\Bigg|^2 \,, \nonumber
\end{align}
with the following definitions of the Passarino-Veltman functions
\begin{eqnarray}
 \frac{i}{16\pi^2} B_0(p^2,m_1^2,m_2^2) &=& \bar \mu^{4-D} \int\!\!\frac{{\rm d}^Dq}{(2\pi)^D} \frac{1}{(q^2-m_1^2)((q+p)^2-m_2^2)} ~, \\
 \frac{i}{16\pi^2} C_0(p^2,k^2,(p+k)^2,m_1^2,m_2^2,m_3^2) &=& \int\!\!\frac{{\rm d}^4q}{(2\pi)^4} \frac{1}{(q^2-m_1^2)((q+p)^2-m_2^2)((q+p+k)^2-m_3^2)} ~.
\end{eqnarray}
The derivatives in Eq.~\eqref{eq:F} act on the last argument of the functions, i.e.
\begin{eqnarray}
 B^\prime_0(a,b,c) &=& \frac{\partial}{\partial c} B_0(a,b,c) ~, \\
 C^\prime_0(a,b,c,d,e,f) &=& \frac{\partial}{\partial f} C_0(a,b,c,d,e,f) ~.
\end{eqnarray}

\section{Higgs Signal Strength Fit}\label{app:fit}

Away from the alignment limit $\cos(\beta-\alpha) = 0$, the couplings of the 125\,GeV Higgs boson differ from their SM predictions.
Therefore, signal strength measurements from ATLAS and CMS can be used to constrain the parameter space of our 2HDMs.
With respect to our previous signal strength analysis in~\cite{Altmannshofer:2018bch}, we include LHC Run 2 updates of $h \to WW$~\cite{Aaboud:2018jqu,Sirunyan:2018egh,Aad:2019lpq}, $h \to \tau\tau$~\cite{Aaboud:2018pen,Sirunyan:2018cpi}, $h \to \mu\mu$~\cite{Sirunyan:2018hbu}, the recent $h \to bb$ observations~\cite{Aaboud:2018zhk,Sirunyan:2018kst} and the results for $tth$ production~\cite{Sirunyan:2018hoz,Aaboud:2018urx}.

\begin{figure}[tb]
\begin{center}
\includegraphics[width=0.46\textwidth]{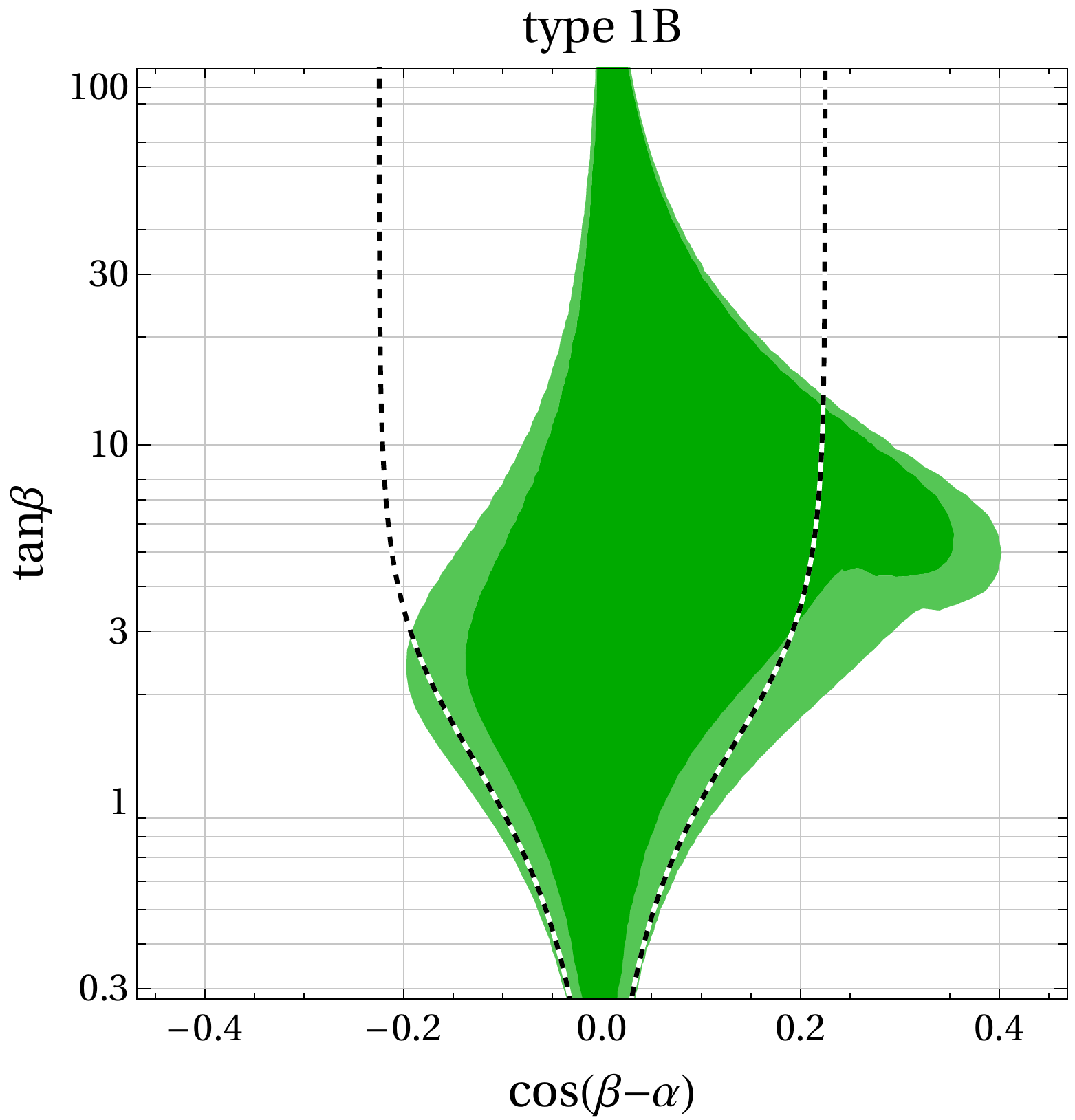}~~~~~~
\includegraphics[width=0.46\textwidth]{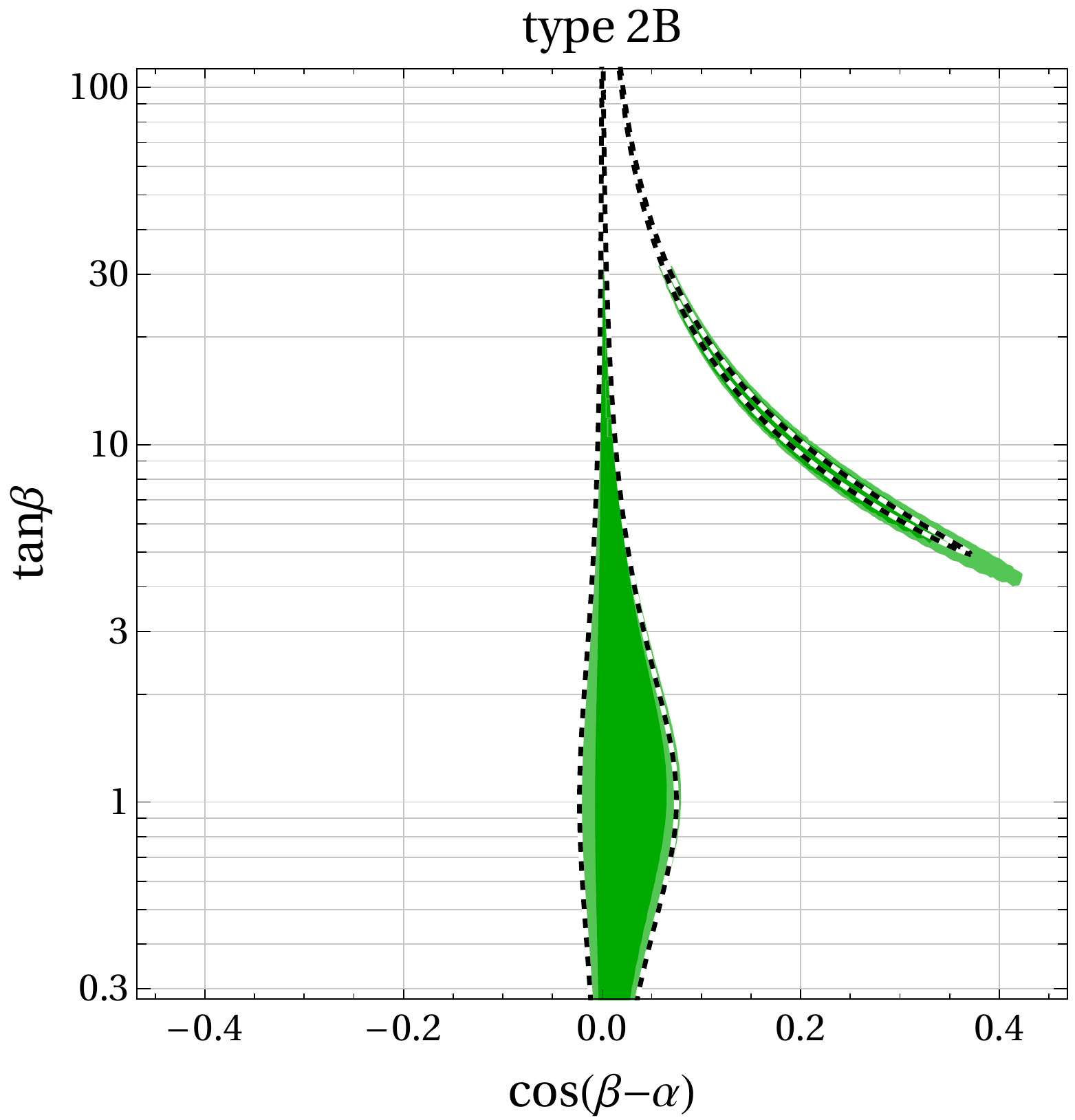} \\[12pt]
\includegraphics[width=0.46\textwidth]{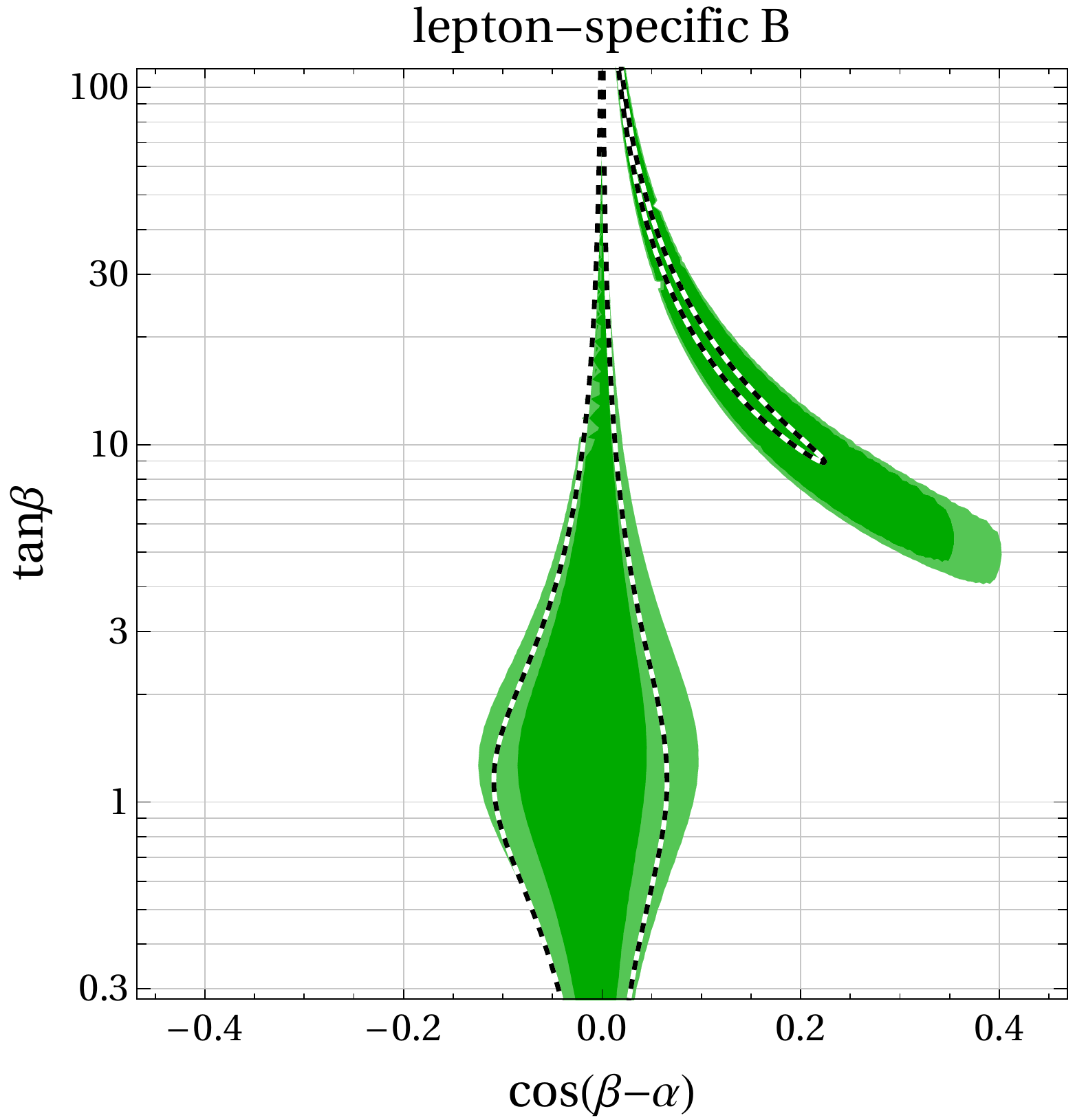}~~~~~~
\includegraphics[width=0.46\textwidth]{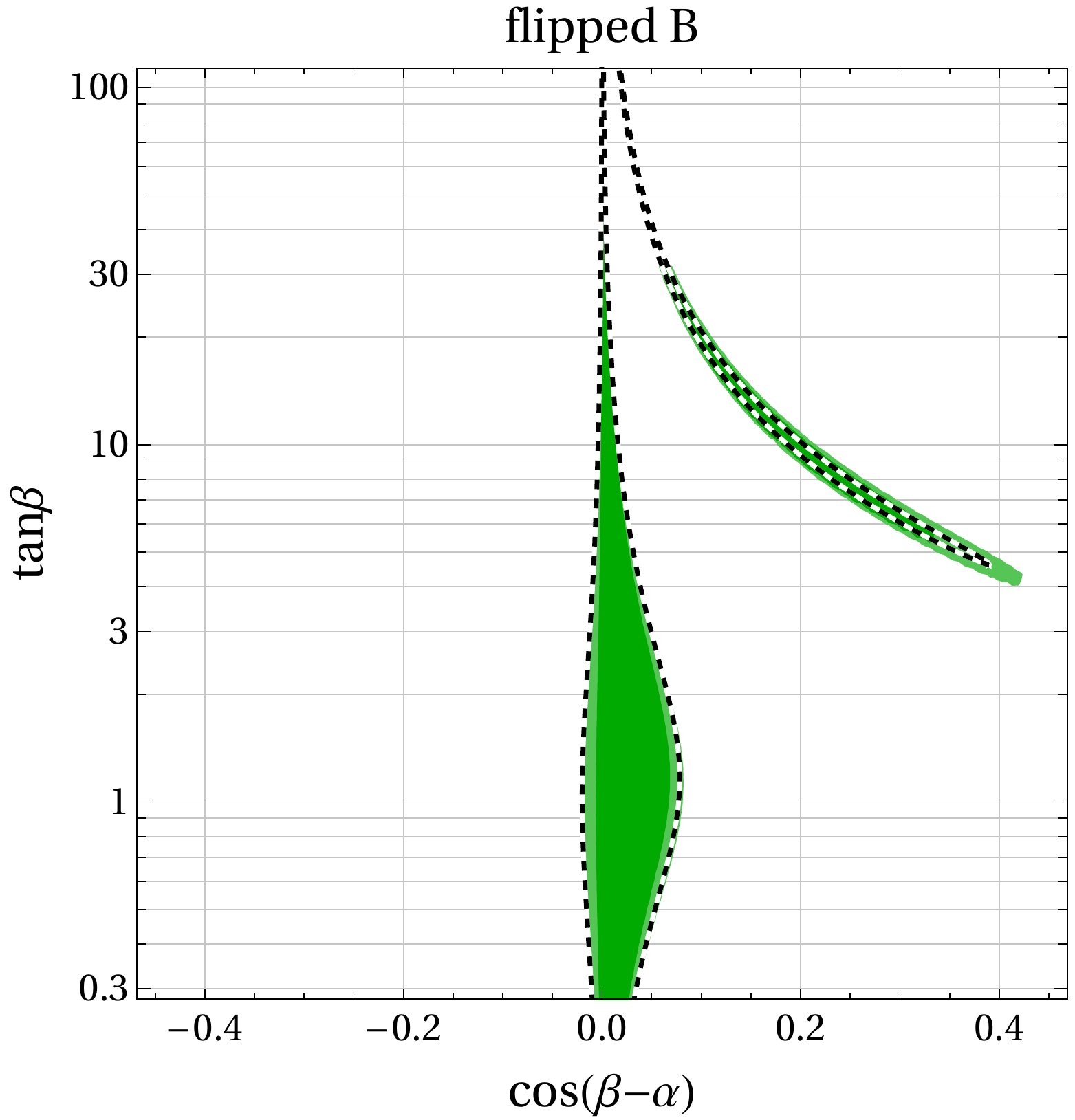}
\caption{Constraints in the $\cos(\beta-\alpha)$ vs. $\tan\beta$ plane based on LHC measurements of the 125\,GeV Higgs signal strengths. Parameter space of the flavorful 2HDMs that is compatible with the data at the $1\sigma$ and $2\sigma$ level is shown in green. For comparison, the $2\sigma$ regions in the corresponding 2HDMs with natural flavor conservation are shown by dashed contours.}
\label{fig:fit}
\end{center}
\end{figure}

In Fig.~\ref{fig:fit} we show the allowed ranges in the $\cos(\beta-\alpha)$ vs. $\tan\beta$ plane in the type~IB model (top left), type~IIB model (top right), lepton-specific~B model (bottom left), and flipped~B model (bottom right) at $1\sigma$ (dark green) and $2\sigma$ (light green). The dotted lines indicate the $2\sigma$ constraint in the corresponding 2HDMs with natural flavor conservation.

\small

\end{document}